\documentclass[aps,showpacs,preprintnumbers,nofootinbib,amsmath,amssymb]{revtex4}
\usepackage[dvips]{epsfig}

\newcommand{\nc}[1]{\newcommand{#1}}

\nc{\its}[1]{\itshape #1 \upshape}
\nc{\mc}[3]{\multicolumn{#1}{#2}{#3}}

\nc{\bc}{\begin{center}}
\nc{\ec}{\end{center}}
\nc{\ig}[1]{\bc \includegraphics{#1} \ec}

\nc{\bo}[1]{\mbox{\boldmath \( #1 \! \! \)  \unboldmath}}
\newcommand{\beqn} {\begin{equation}}
\newcommand{\eqn} {\end{equation}}
\nc{\be}{\begin{eqnarray}}
\nc{\ee}{\end{eqnarray}}
\nc{\bew}{\begin{eqnarray*}}
\nc{\eew}{\end{eqnarray*}}
\nc{\bs}{\begin{subeqnarray}}   
\nc{\es}{\end{subeqnarray}}     
\nc{\nnn}{\nonumber \\}
\nc{\f}[2]{\frac{#1}{#2}}
\nc{\td}[2]{\f{d #1}{d #2}}
\nc{\pd}[2]{\f{\partial #1}{\partial #2}}
\nc{\suli}{\sum\limits}
\nc{\proli}{\prod\limits}
\nc{\ili}{\int\limits}
\nc{\sr}[2]{\stackrel{#1}{#2}}
\nc{\dps}{\displaystyle}
\nc{\ket}[1]{\left| #1 \right>}
\nc{\bra}[1]{\left< #1 \right|}
\nc{\bracket}[2]{\left< #1 \right| \left. \! #2 \right>}
\nc{\norm}[1]{\left\| #1 \right\|}
\nc{\lndm}[1]{\pd{^{#1} \ln{\det{M}}}{\mu^{#1}}}
\nc{\pdmm}[1]{M^{-1} \pd{^{#1} M}{\mu^{#1}}}
\nc{\pdm}{M^{-1}\pd{M}{\mu}}
\nc{\trac}[1]{\mbox{Tr}\left(#1\right)}
\nc{\hm}{\hat{m}}
\nc{\hr}{\hat{r}}

\def\lsim{\raise0.3ex\hbox{$<$\kern-0.75em\raise-1.1ex\hbox{$\sim$}}}
\def\gsim{\raise0.3ex\hbox{$>$\kern-0.75em\raise-1.1ex\hbox{$\sim$}}}

\begin{document}

\title{The QCD Equation of State with almost Physical Quark Masses}

\author{M. Cheng$^{\rm a}$, N. H. Christ$^{\rm a}$, S. Datta$^{\rm b}$, 
J. van der Heide$^{\rm c}$,
C. Jung$^{\rm d}$, F. Karsch$^{\rm c,d}$, O. Kaczmarek$^{\rm c}$,\\ 
E. Laermann$^{\rm c}$, R. D. Mawhinney$^{\rm a}$, C. Miao$^{\rm c}$,
P. Petreczky$^{\rm d,e}$, K. Petrov$^{\rm f}$,
C. Schmidt$^{\rm d}$, W. Soeldner$^{\rm d}$ and T. Umeda$^{\rm g}$
}

\affiliation{
$^{\rm a}$ Physics Department,Columbia University, New York, NY 10027, USA\\
$^{\rm b}$ Department of Theoretical Physics, Tata Institute of Fundamental
Research, Homi Bhabha Road, Mumbai 400005, India\\
$^{\rm c}$Fakult\"at f\"ur Physik, Universit\"at Bielefeld, D-33615 Bielefeld,
Germany\\
$^{\rm d}$Physics Department, Brookhaven National Laboratory, 
Upton, NY 11973, USA \\
$^{\rm e}$ RIKEN-BNL Research Center, Brookhaven National Laboratory, 
Upton, NY 11973, USA \\
$^{\rm f}$Niels Bohr Institute, University of Copenhagen, 
Blegdamsvej 17, DK-2100 Copenhagen, Denmark\\
$^{\rm g}$Graduate School of Pure and Applied Sciences, University of Tsukuba,
Tsukuba, Ibaraki 305-8571, Japan\\
}

\date{\today}
\preprint{BNL-NT-07/38}
\preprint{BI-TP 2007/20}
\preprint{CU-TP-1181}

\begin{abstract}
We present results on the equation of state in QCD with
two light quark flavors and a heavier strange quark. 
Calculations with improved staggered fermions have been performed on 
lattices with temporal extent $N_\tau =4$ and $6$ on a line of constant 
physics with almost physical quark mass values; the pion mass is about 
$220$~MeV, and the strange quark mass is adjusted to its physical value.
High statistics results on large lattices are obtained for bulk 
thermodynamic observables, {\it i.e.} pressure, energy and entropy density,
at vanishing quark chemical potential for a wide range 
of temperatures, $140$~MeV$\; \le \; T\; \le 800$~MeV.
We present a detailed
discussion of finite cut-off effects which become particularly significant
for temperatures larger than about twice the transition temperature. At these
high temperatures we also performed calculations of the trace anomaly on
lattices with temporal extent $N_\tau=8$.
Furthermore, we have performed an extensive analysis of zero temperature
observables including the light and strange quark condensates and
the static quark potential at zero temperature. These are used to set the
temperature scale for thermodynamic observables and to calculate renormalized
observables that are sensitive to deconfinement and chiral symmetry restoration
and become order parameters in the infinite and zero quark mass limits, respectively. 
\end{abstract}


\maketitle

\section{Introduction}
\label{intro}

Reaching a detailed understanding of bulk thermodynamics of QCD, e.g. the
temperature dependence of pressure and energy density as well as the
equation of state, $p(\epsilon)$ vs. $\epsilon$, is one of the central
goals of non-perturbative studies of QCD on the lattice. 
The equation of state clearly is of central importance for the
understanding of thermal properties of any thermodynamic system. It 
provides direct insight into the relevant degrees of freedom and
their correlation in different phases of strongly interacting matter.  
We have some understanding of the equation of state in limiting 
cases of high and low temperatures from perturbation 
theory \cite{Kap79,Zhai,Arnold,Vuorinen,Kajantie}
and hadron gas phenomenology \cite{redlich}, respectively. 
In the transition region from the low temperature hadronic regime to
the high temperature quark gluon plasma, however, one has to
rely on a genuine non-perturbative approach, lattice regularized QCD, to
study the non-perturbative properties of strongly interacting matter.

Lattice studies of bulk thermodynamics are particularly demanding as the
most interesting observables, pressure and energy density, are given in 
terms of differences of dimension 4 operators. These differences
are particularly difficult to evaluate because
both terms being subtracted contain the pressure or energy
density of the vacuum, an unphysical quantity that is approximately 
$1/(aT)^4$ larger than the sought-after difference.  
Numerical signals thus rapidly decrease with the fourth power of the lattice
spacing, $a$, when one tries to approach the continuum limit at fixed 
temperature ($T$). For this reason improved
actions, which allow one to perform calculations on rather coarse lattices with 
relatively small lattice discretization errors, are quite useful in 
thermodynamic calculations. Indeed, the early calculations of bulk 
thermodynamics with standard staggered \cite{Bernard} and Wilson 
\cite{AliKhan} fermion discretization schemes have shown that at high 
temperature
bulk thermodynamic observables are particularly sensitive to lattice 
discretization errors. This closely follows observations made in studies
of the thermodynamics of SU(3) gauge theories \cite{Boyd}. 
In order to minimize discretization errors at high temperature, improved
staggered fermion actions - the p4-action \cite{Peikert} and the asqtad
action \cite{milc_eos} - have been used to study the QCD equation of state. 
Recent studies, performed with the asqtad action with almost physical 
quark mass values on lattices with two different values of the lattice 
cut-off \cite{milc_eos}, indeed show much smaller discretization errors 
than similar studies performed with the 1-link, stout smeared staggered 
fermion action \cite{aoki}. 
Another source for cut-off errors arises, however, from the explicit breaking
of flavor symmetry in the staggered fermion formulation. While this is not
of much concern in the chirally symmetric high temperature phase of QCD,
it leads to cut-off dependent modifications of the hadron spectrum and thus 
may influence the calculation of thermodynamic observables in the low 
temperature hadronic phase of QCD. Techniques to reduce flavor symmetry
breaking through the introduction of so-called 'fat links' are thus 
generally exploited in numerical calculations with staggered 
fermions \cite{fat}.   

In this paper we report on a calculation of bulk thermodynamics  
in QCD with almost physical light quark masses and a physical
value of the strange quark mass. Our calculations have been performed with 
a tree level Symanzik-improved gauge action and an improved staggered fermion 
action, the p4-action with 3-link smearing (p4fat3), which 
removes ${\cal O}(a^2)$ cut-off effects  at
tree-level and also leads to small cut-off effects in ${\cal O}(g^2)$ 
perturbation theory \cite{Heller}.  
At each temperature, we perform simulations with two degenerate light quark 
masses and a heavier 
strange quark mass for two different values of the lattice cut-off, 
corresponding to lattices with temporal extent  $N_\tau =4$ and $6$. 
In these calculations we explore a wide range of temperatures varying 
from about $140$~MeV to about $800$~MeV.
This corresponds to the temperature interval relevant for current 
experimental studies of dense matter in heavy ion collisions at RHIC as 
well as the forthcoming experiments at the LHC.
Bare quark masses have been adjusted to keep physical masses
approximately constant when the lattice-cut off is varied. 
At high temperatures, $T\gsim 350$~MeV, we also performed calculations
on lattices with temporal extent $N_\tau =8$ to get control over 
cut-off effects in the high temperature limit.

We will start
in the next section by reviewing basic thermodynamic relations in the 
continuum valid for thermodynamic calculations on such {\it lines of 
constant physics} (LCP). In section III we outline
details of our calculational set-up with improved staggered fermions. 
In section IV we present our zero temperature calculations needed
to define the line of constant physics and the temperature scale
deduced from properties of the static quark-antiquark potential.
Section V is devoted to the presentation of our basic result, the difference
between energy density and three times the pressure from which we
obtain all other thermodynamic observables, e.g. the pressure,
energy and entropy densities as well as the velocity of sound. 
Section VI is devoted to a discussion of the temperature
dependence of  Polyakov loop expectation values and chiral condensates
which provides a comparison between the deconfining and chiral symmetry
restoring features of the QCD transition. We finally present a discussion
of our results and a comparison with other improved staggered fermion 
calculations of bulk thermodynamics in Section VII.

\section{Thermodynamics on lines of constant physics}
\label{LCP}

To start our discussion of QCD thermodynamics on the lattice we recall
some basic thermodynamic relations in the continuum. For large, homogeneous
media the basic bulk thermodynamic observables we will consider here can
be obtained from the grand canonical partition function with vanishing
quark chemical potentials, $Z(T,V)$. We introduce the
grand canonical potential, $\Omega(T,V)$, normalized such that it vanishes at 
vanishing temperature, 
\beqn
\Omega(T,V) = T\; {\rm ln} Z(T,V) -\Omega_0\;\; ,
\label{gcpotential}
\eqn
with $\displaystyle{\Omega_0 =\lim_{T\rightarrow 0} T\; {\rm ln} Z(T,V)}$.
With this we obtain the thermal part of the pressure ($p$) 
and energy density ($\epsilon$)
\beqn
p  =  \frac{1}{V} \Omega(T,V)  \;\; ,\;\; 
\epsilon = \frac{T^2}{V} \frac{\partial \Omega(T,V) / T}{\partial T}~~,~~ 
\label{freeenergy}
\eqn
which by construction both vanish at vanishing temperature.
Using these relations one can express the difference between 
$\epsilon$ and $3p$, {\it i.e} the thermal contribution to the trace of 
the energy-momentum tensor $\Theta^{\mu\mu} (T)$, in terms of a derivative 
of the pressure with 
respect to temperature, {\it i.e.}
\begin{eqnarray}
\frac{\Theta^{\mu\mu} (T)}{T^4} \equiv \frac{\epsilon - 3p }{T^4}  =  
T \frac{\partial}{\partial T} (p/T^4) ~~,
\label{delta}
\end{eqnarray}

In fact, it is $\Theta^{\mu\mu} (T)$ which is the
basic thermodynamic quantity conveniently calculated on the lattice. 
All other bulk thermodynamic observables, e.g. 
$p/T^4$, $\epsilon/T^4$ as well as the entropy density, $s/T^3\equiv
(\epsilon+p)/T^4$, can be deduced from this using the above thermodynamic
relations. 
In particular, we obtain the pressure from $\Theta^{\mu\mu} (T)$
through integration of Eq.~\ref{delta},
\beqn
\frac{p(T)}{T^4} - \frac{p(T_0)}{T_0^4} = \int_{T_0}^{T} {\rm d}T' 
\frac{1}{T'^5} \Theta^{\mu\mu} (T') \;\; .
\label{pressure}
\eqn
Usually, the temperature for the lower integration limit, $T_0$, is chosen to 
be a temperature sufficiently deep in the hadronic phase of QCD where the 
pressure $p(T_0)$, receives contributions only from massive hadronic states 
and is already exponentially small. We will discuss this in more detail in 
Section V.
Eq.~\ref{pressure} then directly gives the pressure at temperature $T$.
Using $p/T^4$ determined from Eq.~\ref{pressure} and combining it with 
Eq.~\ref{delta}, we obtain $\epsilon/T^4$ as well as $s/T^3$. This makes it
evident, that there is indeed only one independent bulk thermodynamic 
observable calculated in the thermodynamic (large volume) limit. 
All other observables are derived through standard thermodynamic relations 
so that thermodynamic consistency of all bulk thermodynamic observables is 
insured by construction!

We stress that the normalization introduced here for the grand 
canonical potential, Eq.~\ref{gcpotential}, forces the pressure and energy 
density to vanish at $T=0$. As a consequence of this normalization, any 
non-perturbative structure of the QCD vacuum, e.g.
quark and gluon condensates, that contribute to the trace anomaly 
$\Theta^{\mu\mu}(0)$, and would lead to a non-vanishing vacuum 
pressure and/or energy density, eventually will show up as non-perturbative 
contributions to the high temperature part of these thermodynamic observables.
This is similar to the normalization used, e.g. in the bag model and the
hadron resonance gas, but differs from the normalization used e.g.
in resummed perturbative calculation at high 
temperature \cite{Blaizot,Andersen} or phenomenological 
(quasi-particle) models for the high temperature phase of QCD \cite{quasi}. 
This should be kept in mind when comparing results for the EoS with 
perturbative and model calculations.  We also note that ambiguities in 
normalizing pressure and energy density at zero temperature drop out in 
a calculation of the entropy density which thus is the preferred observable
for such comparisons.

\section{Lattice Formulation}
\label{setup}

In a lattice calculation, temperature and volume are given in terms of the 
temporal ($N_\tau$) and spatial ($N_\sigma$) lattice extent as well as the 
lattice spacing, $a$, which is controlled through the lattice gauge coupling 
$\beta \equiv 6/g^2$, 
\beqn
T=\frac{1}{N_\tau a(\beta)} \;\; , \;\; V = \left( N_\sigma a(\beta) \right)^3
\; .
\label{temperature}
\eqn
As all observables that are calculated on the lattice, are functions of the 
coupling, $\beta$, we may rewrite Eq.~\ref{delta} in terms
of a derivative taken with respect to $\beta$ rather than $T$. Furthermore
we adopt the normalization of the pressure as introduced in 
Eq.~\ref{gcpotential}. This
insures a proper renormalization of thermodynamic quantities and, as a 
consequence, forces the pressure to vanish in the vacuum, {\it i.e.} at $T=0$.

Let us write the QCD partition function on a lattice of size 
$N_\sigma^3 N_\tau$ as
\beqn
Z_{LCP}(\beta, N_\sigma, N_\tau) =  \int \prod_{x,\mu} {\rm d}U_{x,\mu}
 {\rm e}^{- S(U)} \;\; ,
\label{partition}
\eqn
where $U_{x,\mu}\in SU(3)$ denotes the gauge link variables and
$S(U) = \beta S_G(U) - S_F(U,\beta)$ is the Euclidean action, 
which is composed out of a purely gluonic contribution, $S_G(U)$, and the
fermionic part, $S_F(U,\beta)$, which arises after integration over the 
fermion fields.
We  will specify this action in more detail in the next section but note here
that we will use tree level improved gauge and fermion actions. 
Although it would be straightforward to introduce one-loop or tadpole
improvement factors in the action the setup used here greatly simplifies
the analysis of thermodynamic observables and in some cases also gives a more 
direct relation to corresponding observables in the continuum. 

When using only tree level improvement the gluonic 
action does not depend on the gauge coupling, $\beta$, and 
the fermion action depends on $\beta$ only through the bare light $(\hm_l)$  
and strange $(\hm_s)$ quark masses. 
The subscript $LCP$ in Eq.~\ref{partition} indicates that we 
have defined the partition function $Z_{LCP}$ on a line of constant
physics (LCP), {\it i.e.} when approaching the continuum limit by increasing
the gauge coupling $(\beta \rightarrow \infty)$ the bare quark 
masses $(\hm_l(\beta),\; \hm_s(\beta))$ in the QCD Lagrangian are tuned 
towards zero such that the vacuum properties of QCD remain unchanged. 
The quark masses thus are not independent parameters
but are functions of $\beta$ which are determined through constraints imposed 
on zero temperature observables; e.g. one demands that a set of hadron masses
remains unchanged when the continuum limit is approached on a LCP.

We now may rewrite Eq.~\ref{delta} in terms of observables calculable
in lattice calculations at zero and non-zero temperature,
\begin{eqnarray}
\frac{\Theta^{\mu\mu} (T)}{T^4}
&=&  -R_\beta(\beta) N_\tau^4 \left( 
\frac{1}{N_\sigma^3 N_\tau}\left\langle \frac{{\rm d}S}{{\rm d}\beta}\right
\rangle_\tau - 
\frac{1}{N_\sigma^3 N_0}\left\langle \frac{{\rm d}S}{{\rm d}\beta}\right
\rangle_0 
\right)
\;\; .
\label{deltaLGT}
\end{eqnarray}
Here $\langle ... \rangle_x$, with $x=\tau,~0$ denote expectation 
values evaluated on finite temperature lattices of size $N_\sigma^3 N_\tau$,
with $N_\tau \ll N_\sigma$,  
and zero temperature lattices, {\it i.e.} on lattices with large temporal 
extent, $N_\sigma^3 N_\tau$ with $N_\tau\equiv N_0 \gsim N_\sigma$, 
respectively. 
Furthermore, $R_\beta$ denotes the lattice version of the QCD $\beta$-function
which arises as a multiplicative factor in the definition of 
$\Theta^{\mu\mu} (T)$
because derivatives with respect to $T$ have been converted to derivatives
with respect to the lattice spacing $a$ on lattices with fixed temporal
extent $N_\tau$,
\beqn
R_{\beta}(\beta) \equiv T{{\rm d}\beta \over {\rm d} T} = -
a{{\rm d}\beta \over {\rm d}a} ~~~.
\label{rgeqn}
\eqn

We note that in the weak coupling, large $\beta$ limit, $R_\beta$ approaches 
the universal form of the 2-loop $\beta$-function of 3-flavor QCD,
\begin{eqnarray}
R_\beta(\beta)
= 12 b_0 + 72 b_1/\beta +{\cal O}(\beta^{-2}) \;\; , 
\label{betaasym} 
\end{eqnarray}
with $b_0 = 9/16\pi^2$ and $b_1=1/4\pi^4$.

We analyze the thermodynamics of QCD with two degenerate light quarks 
($\hm_l\equiv \hm_u= \hm_d$) and a heavier strange quark ($\hm_s$)
described by the QCD partition function given in Eq.~\ref{partition}. 
For our studies of bulk thermodynamics we use the same discretization
scheme which has been used recently by us in the study
of the QCD transition temperature \cite{p4_Tc}, {\it i.e.}   
we use an ${\cal O}(a^2)$ tree level improved gauge action constructed from 
a 4-link plaquette term and a planar 6-link Wilson loop as well as a 
staggered fermion action that contains a smeared 1-link term and bent
3-link terms. We call this action the p4fat3-action; further details are 
given in Ref.~\cite{Peikert} where the p4fat3 action was first
used in studies of the QCD equation of state on lattices
with temporal extent $N_\tau=4$ and larger quark masses.  
With this action, bulk thermodynamic
quantities like pressure and energy density are ${\cal O}(a^2)$
tree level improved; corrections to the high temperature ideal
gas limit only start at ${\cal O}(1/N_\tau^{4})$ and are significantly
smaller than for the Naik action or the standard staggered action which
suffers from large ${\cal O}(1/N_\tau^2)$ cut-off effects at high temperature. 
An analysis of cut-off
effects in the ideal gas limit and in ${\cal O}(g^2)$ lattice perturbation
theory \cite{Heller} shows that deviations from perturbative results 
are already only a few percent for lattices with temporal extent
as small as $N_\tau =6$.

Following the notation used in Ref.~\cite{p4_Tc} the Euclidean action is 
given as 
\beqn
S(U) = \beta S_G(U) - S_F(U,\beta)\;\; ,
\label{action}
\eqn
with a gluonic contribution, $S_G(U)$, and a fermionic 
part, $S_F(U,\beta)$. The latter can be expressed in terms of 
the staggered fermion matrices, $D_{\hm_l (\hm_s)}$, for two 
light ($\hm_l$) and a heavier strange quark ($\hm_s$),
\begin{eqnarray}
S_F(U,\beta) &=& \frac{1}{2} {\rm Tr}\ln  D(\hat{m}_l(\beta))
+ \frac{1}{4} {\rm Tr}\ln  D(\hat{m}_s(\beta)) \;\; .
\label{SF}
\end{eqnarray}
Here we took the fourth root of the staggered fermion determinant to
represent the contribution of a single fermion flavor to the QCD 
partition function\footnote{There is a controversy regarding the 
validity of the rooting approach in numerical calculations with
staggered fermions. For further details we refer to recent reviews presented
at Lattice conference  \protect\cite{sharpe,creutz,kronfeld} and references 
therein.}. 

We also introduce the light and strange quark condensates calculated
at finite ($x=\tau$) and zero temperature ($x=0$), respectively,
\beqn
\langle\bar{\psi}\psi\rangle_{q,x} \equiv \frac{1}{4}
\frac{1}{N_\sigma^3N_x} \left\langle {\rm Tr} D^{-1}(\hat{m}_q) 
\right\rangle_x \;\; ,\;\; 
q=l,~s\;\; , \;\; x=0,~\tau \;\; ,
\label{quarkcondensate}
\eqn
as well as expectation values of the gluonic action density,
\beqn
\langle s_G\rangle_x \equiv \frac{1}{N_\sigma^3N_x} \left\langle S_G
\right\rangle_x \;\; .
\label{gluondensity}
\eqn
All numerical calculations have been performed using the Rational Hybrid Monte
Carlo (RHMC) algorithm \cite{rhmc}
with parameters that have been optimized \cite{p4_Tc} to reach acceptance
rates of about 70\%. Some details on our tuning of parameters of the
RHMC algorithm have been given in \cite{p4_Tc_nf3}.

For the discussion of the thermodynamics on a line of 
constant physics (LCP) it sometimes is convenient to parametrize the quark mass
dependence of $S_F$ in terms of the light quark mass $\hm_l$ and 
the ratio $h\equiv \hm_s/\hm_l$ rather than $\hm_l$ and $\hm_s$ separately.
We thus write the $\beta$-dependence of the strange quark mass as,
$\hm_s(\beta) = \hm_l (\beta) h(\beta)$. 
In the evaluation of $(\epsilon -3p)/T^4$
we then will need to know the derivatives of these parametrizations 
with respect to $\beta$. We define 
\begin{eqnarray}
R_m(\beta) = \frac{1}{\hm_l(\beta)}
\frac{{\rm d} \hm_l(\beta)}{{\rm d}\beta} &,&
R_h(\beta) = \frac{1}{h(\beta)} \frac{{\rm d} h(\beta)}{{\rm d}\beta} \; . 
\label{massfunctions}
\end{eqnarray}
With these definitions we may rewrite Eq.~\ref{deltaLGT} as
\begin{eqnarray}
\frac{\epsilon-3p}{T^4}
&=& T\frac{{\rm d}}{{\rm d}T} \left( \frac{p}{T^4}\right)
= R_\beta (\beta) \frac{\partial p/T^4}{\partial \beta}\nonumber \\
&=&
\frac{\Theta^{\mu\mu}_G(T)}{T^4} +
\frac{\Theta^{\mu\mu}_F(T)}{T^4} +
\frac{\Theta^{\mu\mu}_h(T)}{T^4}  \;\; ,
\label{e3p}
\end{eqnarray}
with
\begin{eqnarray}
\frac{\Theta^{\mu\mu}_G(T)}{T^4}  &=& 
R_\beta
\left[ \langle s_G \rangle_0 - \langle s_G \rangle_\tau \right] N_\tau^4 \; , 
\label{e3pgluon}  \\
\frac{\Theta^{\mu\mu}_F(T)}{T^4}  
&=& - R_\beta R_{m}
\left[ 2 \hm_l
\left(\langle\bar{\psi}\psi \rangle_{l,0}
- \langle\bar{\psi}\psi \rangle_{l,\tau}\right)
+ \hm_s \left(\langle\bar{\psi}\psi \rangle_{s,0}
- \langle\bar{\psi}\psi \rangle_{s,\tau}\right)
\right] N_\tau^4 \; , \label{e3pfermion}\\
\frac{\Theta^{\mu\mu}_h(T)}{T^4}  &=& 
- R_\beta R_{h} \hm_s \left[ \langle\bar{\psi}\psi \rangle_{s,0}
- \langle\bar{\psi}\psi \rangle_{s,\tau}\right] N_\tau^4 \; . 
\label{e3prest}
\end{eqnarray}
We will show in the next section that to a good approximation $h(\beta)$
stays constant on a LCP. $R_h$ thus vanishes on the LCP and 
consequently the last term in Eq.~\ref{e3p}, $\Theta^{\mu\mu}_h$, will 
not contribute to the thermal part of the trace anomaly, $\Theta^{\mu\mu} (T)$. 
The other two terms stay finite in the continuum limit and correspond 
to the contribution of the thermal parts of gluon and quark condensates
to the trace anomaly.
We note that the latter contribution vanishes in the chiral limit of three 
flavor QCD ($\hm_l,\; \hm_s\rightarrow 0$). The trace anomaly would then
entirely be given by $\Theta^{\mu\mu}_G (T)$ and the observables  
entering the calculation of bulk thermodynamic quantities in the chiral limit 
of QCD would reduce to those needed also in a pure $SU(3)$ gauge theory
\cite{Boyd}. In fact, we also find that for physical values of the quark
masses the trace anomaly is dominated by the gluonic contribution, 
$\Theta^{\mu\mu}_G (T)$. As will become clear in Section V 
$\Theta_F^{\mu \mu}(T)$ contributes less than 10\% to the total trace anomaly 
for temperatures large than about twice the transition temperature.

We also note that the prefactor in Eq.~\ref{e3pfermion} will approach unity
in the continuum limit as $R_m$ attains a universal form up to 2-loop level
which is similar to that of $R_\beta^{-1}$ and is only modified through the
anomalous dimension of the quark mass renormalization \cite{Gasser}. For
the relevant combination of $\beta$-functions that enters the fermionic part
of the trace anomaly, one has
\begin{equation}
-\left( R_\beta(\beta) R_m(\beta)\right)^{1-loop} = 
1 +\frac{16b_0}{3\beta}\;\; .
\label{Rm2loop}
\end{equation}

\section{Static quark potential and the line of constant physics}

\subsection{Construction of the line of constant physics}

We will calculate thermodynamic observables on a line of constant physics (LCP)
that is defined at $T=0$ as a line in the space of light and 
strange bare quark 
masses parametrized by the gauge coupling $\beta$. Each point on this
line corresponds to identical physical conditions at different values
of the lattice cut-off which is tuned towards the continuum limit
by increasing the gauge coupling $\beta$. We define the line of
constant physics by demanding (i) that the ratio of masses for the strange 
pseudo-scalar and the kaon mass, $m_{\bar{s}s}/ m_K$, stays constant
and (ii) that $m_{\bar{s}s}$
expressed in units of the scale parameter $r_0$ stays constant. 
The latter gives the distance at which the slope of the zero temperature,
static quark potential, $V_{\bar{q}q}(r)$, attains a certain value. We 
also introduce the scale $r_1$,
which frequently is used on finer lattices to convert lattice results 
expressed in units of the cut-off to physical scales,
\beqn
\left( r^2 \frac{{\rm d}V_{\bar{q}q}(r)}{{\rm d}r} \right)_{r=r_0} = 
1.65 \;\; , \;\; 
\left( r^2 \frac{{\rm d}V_{\bar{q}q}(r)}{{\rm d}r} \right)_{r=r_1} =
1.0  \;\; . 
\label{potential}
\eqn 
We checked that (i) and (ii)  also hold true, if we replace $m_{\bar{s}s}$
by the mass of the light quark pseudo-scalar meson, $m_\pi$. 
However, errors on $m_\pi r_0$ and $(m_\pi/ m_K)$ are generally larger 
which, in particular at large values of $\beta$ makes the parametrization 
of the LCP less stringent.  

Leading order chiral perturbation theory suggests that the ratio 
$(m_{\bar{s}s}/ m_K)^2$ is proportional to $\hm_s/(\hm_l+\hm_s)$.
One thus expects this ratio to stay constant for fixed $h=\hm_s/\hm_l$.
This is, indeed fulfilled in the entire regime of couplings, $\beta$,
explored in our calculations (see Table~\ref{tab:masses_r0}). 
The first condition for fixing the LCP parameters thus, in practice, has 
been replaced by choosing $h=\hm_l / \hm_s$ to be constant.  
As a consequence we find $R_h(\beta) = 0$, which simplifies
the calculation of thermodynamic quantities.

In order to define a line of constant strange quark mass, as a second
condition for the LCP we demand that the product  $m_{\bar{s}s}r_0$
stays constant. For our LCP we chose 1.59 as the value for the product.
Here one should note that $m_{\bar{s}s}$ determined in our calculations
only receives contributions from connected diagrams and does not include
disconnected loops.  In order to compare our value (1.59, see discussion
below) to a physical one, we therefore follow the argumentation of
Ref.~\cite{milc_masses} and adopt
$m_{\bar{s}s}=\sqrt{2m_K^2-m_\pi^2}=686$~MeV as the physical mass of our
strange pseudoscalar. Together with the scale $r_0=0.469(7)$~fm as determined
in Ref.~\cite{gray} through a comparison of $r_0$ with level splittings of
the charmonium system \cite{gold}, this yields $m_{\bar{s}s}r_0 \simeq 1.63$.
Of course, there is some ambiguity in this choice as current determinations
of $r_0$ differ by about 10\% \cite{gray,Ishikawa}. This introduces some
systematic error in the definition of the {\it physical LCP}. The main reason
for deviation from the physical LCP in the present calculation, however, is
due to the choice of the light quark masses which are about a factor two too
large.

Fixing the light and strange pseudo-scalar masses in units of $r_0$ required
some trial runs for several $\beta$ values. We then used the leading order
chiral perturbation theory ansatz
$m^2_{\bar{s}s} \sim \hm_s$ (or $m_\pi^2 \sim \hm_l$) to
choose $\hm_s$ and $\hm_l\equiv \hm_s/10$ at several values of the
gauge coupling and used a renormalization group inspired interpolation
to determine quark mass values at several other $\beta$ values at which
high statistics simulations have been performed. It turned out that these
values are best fitted by $m_{\bar{s}s}r_0=1.59$. We thus use this value
rather than the value $1.63$ mentioned above, to define our LCP. For all
other simulations we then used the results of these zero temperature
calculations to determine the quark mass values that belong to a line of
constant physics characterized by:
\begin{equation}
{\rm LCP:}\quad
 (i) \; m_{\bar{s}s} r_0=1.59 \;\; , \;\; (ii)\; h\equiv \hm_s/\hm_l=10 
\nonumber
\end{equation}
In general our calculations are thus performed at parameter values
close to the LCP which is defined by the above condition.
The parameters of all our zero temperature calculations performed to
determine the LCP, results for meson masses and parameters of the
static quark potential are summarized in Table~\ref{tab:masses_r0}. 
As can be seen, at our actual simulation points the results for 
$m_{\bar{s}s} r_0$ fluctuate around the mean value by a few percent. 
We also checked the
sensitivity of the meson masses used to determine the LCP to finite volume
effects. At $\beta=3.49$ and $3.54$ we performed calculations on $32^4$
lattices in addition to the $16^3\cdot 32$ lattices. As can bee seen
from Table I results for $m_{\bar{s}s}$ and $m_K$ agree within statistical
errors and volume effects are at most on the level of 2\% for the light
pseudo-scalar.

The LCP is furthermore characterized by $m_\pi/m_K = 0.435(2)$ and 
$m_{\bar{s}s} /m_K = 1.33(1)$. Using $r_0=0.469(7)$~fm
to convert to physical scales we find that on the LCP 
the light and strange pseudo-scalar masses are
$m_\pi\simeq 220(4)$~MeV, $m_{\bar{s}s}\simeq  669(10)$~MeV and the kaon
mass is given by $m_K\simeq 503(6)$~MeV.

\begin{table}[t]
\begin{center}
\begin{tabular}{|c|c|c|l|l|l|l|l|l|}
\hline
$\beta$ & $100\hat{m}_l$ & $N_\sigma^3\cdot N_\tau$ &
$m_\pi a$ &
$m_{\bar{s}s} a$ &
$m_K a$ &
$r_0/a$ &
$\sqrt{\sigma}a$ & $c(g^2) r_0$ \\
\hline
 3.150 &    1.100 & $16^3\cdot 32$        
 &     0.3410( 2)
 &     1.0474( 1)
 &     0.7854( 2)
  & 1.467(72)   & 0.75(18)  
 &       0.97(12)
\\
 3.210 &    1.000 & $16^3\cdot 32$        
 &     0.3262( 1)
 &     0.9988( 1)
 &     0.7496( 1)
  & 1.583(36)   & 0.685(75) 
 &      1.118(68)
\\
 3.240 &    0.900 & $16^3\cdot 32$        
 &     0.3099( 2)
 &     0.9485( 2)
 &     0.7125( 3)
  & 1.669(31)   & 0.658(36) 
 &      1.243(67)
\\
 3.277 &    0.765 & $16^3\cdot 32$        
 &     0.2881( 7)
 &     0.8769( 5)
 &     0.6599( 6)
  & 1.797(19)   & 0.612(53) 
 &      1.362(53)
\\
 3.290 &    0.650 & $16^3\cdot 32$        
 &     0.2667( 8)
 &     0.8104( 7)
 &     0.6101( 8)
  & 1.823(16)   & 0.623(32) 
 &      1.362(29)
\\
 3.335 &    0.620 & $16^3\cdot 32$        
 &     0.2594( 3)
 &     0.7884( 2)
 &     0.5941( 5)
  & 1.995(11)   & 0.5668(73)
 &      1.504(22)
\\
 3.351 &    0.591 & $16^3\cdot 32$        
 &     0.2541( 7)
 &     0.7692( 5)
 &     0.5800( 7)
  & 2.069(12)   & 0.551(11) 
 &      1.594(24)
\\
 3.382 &    0.520 & $16^3\cdot 32$        
 &     0.2370( 6)
 &     0.7194( 5)
 &     0.5422( 5)
  & 2.230(14)   & 0.5100(82)
 &      1.718(57)
\\
 3.410 &    0.412 & $16^3\cdot 32$        
 &     0.2098( 4)
 &     0.6371( 6)
 &     0.4796( 8)
  & 2.503(18)   & 0.440(10) 
 &      2.073(49)
\\
 3.420 &    0.390 & $24^3\cdot 32$        
 &     0.2029( 8)
 &     0.6177( 5)
 &     0.4675( 5)
  & 2.577(11)   & 0.4313(56)
 &      2.124(33)
\\
 3.430 &    0.370 & $24^3\cdot 32$        
 &     0.1986( 6)
 &     0.6000( 3)
 &     0.4529( 5)
  & 2.6467(81)  & 0.4225(53)
 &      2.178(17)
\\
 3.445 &    0.344 & $24^3\cdot 32$        
 &     0.1909( 7)
 &     0.5749( 4)
 &     0.4335( 5)
  & 2.813(15)   & 0.3951(68)
 &      2.388(35)
\\
 3.455 &    0.329 & $24^3\cdot 32$        
 &     0.1833(10)
 &     0.5580( 6)
 &     0.4204( 8)
  & 2.856(20)   & 0.3895(68)
 &      2.375(42)
\\
 3.460 &    0.313 & $16^3\cdot 32$        
 &     0.1808(16)
 &     0.5443(11)
 &     0.4102(11)
  & 2.890(16)   & 0.3831(84)
 &      2.391(55)
\\
 3.470 &    0.295 & $24^3\cdot 32$        
 &     0.1686(19)
 &     0.5233( 8)
 &     0.3940(12)
  & 3.065(18)   & 0.3592(75)
 &      2.617(41)
\\
 3.490 &    0.290 & $16^3\cdot 32$        
 &     0.1689(14)
 &     0.5115(11)
 &     0.3842(11)
  & 3.223(31)   & 0.3423(66)
 &      2.757(59)
\\
3.490 & 0.290 &   $32^4$&   0.1679(8) & 0.5113(8)& 0.3840(7)&~&~&~
\\
 3.510 &    0.259 & $16^3\cdot 32$        
 &     0.1525(40)
 &     0.4740(20)
 &     0.3554(22)
  & 3.423(61)   & 0.322(14) 
 &      2.934(92)
\\
 3.540 &    0.240 & $16^3\cdot 32$        
 &     0.1495(24)
 &     0.4458(20)
 &     0.3358(19)
  & 3.687(34)   & 0.3011(46)
 &      3.128(51)
\\
3.540 &  0.240  &  $32^4$ &   0.1469(11)& 0.4451(6)& 0.3339(11)&~&~&~
\\
 3.570 &    0.212 & $24^3\cdot 32$        
 &     0.1347(53)
 &     0.4053(18)
 &     0.3028(23)
  & 4.009(26)   & 0.2743(38)
 &      3.414(47)
\\
 3.630 &    0.170 & $24^3\cdot 32$        
 &     0.1126(20)
 &     0.3386( 7)
 &     0.2537( 8)
  & 4.651(41)   & 0.2352(44)
 &      3.939(59)
\\
 3.690 &    0.150 & $24^3\cdot 32$        
 &     0.1020(90)
 &     0.2960(20)
 &     0.2230(30)
  & 5.201(48)   & 0.2116(36)
 &      4.320(63)
\\
 3.760 &    0.130 & $24^2\cdot32\cdot 48$ 
 &     0.0857(32)
 &     0.2530(16)
 &     0.1894(16)
  & 6.050(61)   & 0.1810(29)
 &      4.984(73)
\\
 3.820 &    0.125 & $32^3\cdot 32$        
 &     0.0830(40)
 &     0.2310(38)
 &     0.1744(50)
  & 6.835(44)   & 0.1701(21)
 &       5.541(106)
\\
 3.920 &    0.110 & $32^3\cdot 32$        
 &     0.0750(70)
 &     0.2020(10)
 &     0.1550(20)
  & 7.814(83)   & 0.1423(24)
 &      6.037(72)
\\
 4.080 &    0.081 & $32^3\cdot 32$        
 &     0.0700(70)
 &     0.1567(36)
 &     0.1220(50)
  & 10.39(23)   & 0.1060(35)
 &       7.710(183)
\\
\hline
\end{tabular}
\end{center}
\caption{Light quark and strange pseudo-scalar meson masses and parameters
of the static quark potential calculated on zero temperature lattices of
size $N_\sigma^3 N_\tau$. The last column gives the renormalization constants
times $r_0$ needed to renormalize the heavy quark potential to the string 
potential at distance $r/r_0 = 1.5$.
}
\label{tab:masses_r0}
\end{table}

\subsection{The static quark potential and the scale \boldmath $r_0$}

On the LCP we determine several parameters, e.g. the short distance scale
$r_0$ and the linear slope parameter, the string tension $\sigma$, that 
characterize the shape of the static quark potential calculated at $T=0$ 
in a fixed range of physical distances.
The 
distance $r_0$, defined in Eq.~\ref{potential}, is 
used to define the temperature scale for the thermodynamics calculations.

The static quark potential, $V_{\bar{q}q}(r)$, has been calculated from 
smeared Wilson loops as described in \cite{p4_Tc} for all parameter sets
listed in Table~\ref{tab:masses_r0}. We checked that the
the smeared Wilson loops project well onto the ground
state at all values of the cut-off by verifying the independence of the
extracted potential parameters on the number of smearing levels used in
the analysis.  The set of gauge couplings, $\beta \in [3.15,4.08]$, used in
this analysis
covers a large interval in which the lattice cut-off changes by a factor 6 
from $a\simeq 0.3\;$fm down to $a\simeq 0.05\;$fm. When analyzing the 
static potential over such a wide range of cut-off values one should make
sure that
the potential is analyzed in approximately the same range of physical
distances. The fit interval $[(r/a)_{min},(r/a)_{max}]$ for fits with a
Cornell type ansatz for the static potential thus has been adjusted for the 
different
values of gauge couplings such that it covers approximately the same range
of physical distances, $r_0/2 \; \lsim\; r\; \lsim\; 2 r_0$, or
$0.25 {\rm fm} \lsim r \lsim 1 {\rm fm}$. We confirmed our analysis of the
static quark potential and the determination of $r_0$ also independently by
using spline interpolations which are not biased by a particular ansatz
for the form of the potential. 

\begin{figure}[t]
\begin{center}
\begin{minipage}[c]{17.5cm}
\begin{center}
\epsfig{file=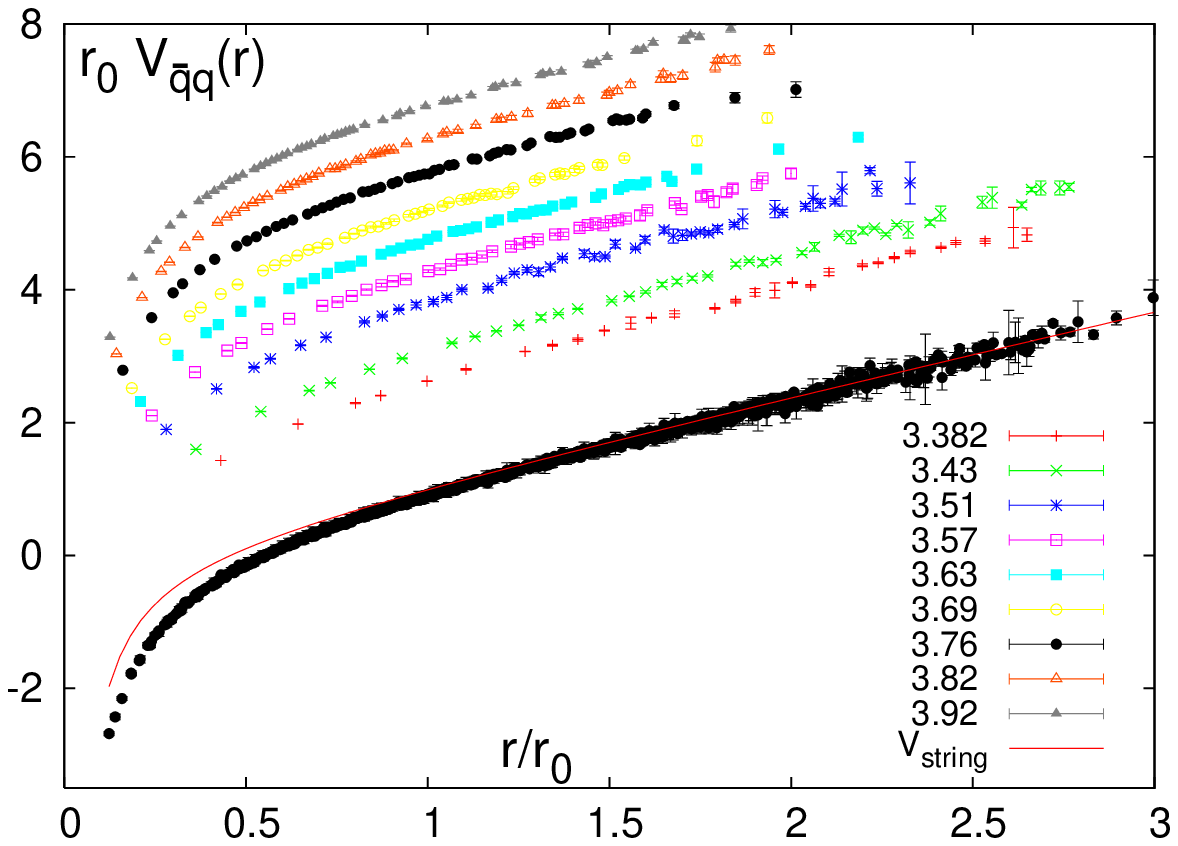, width=8.5cm}
\epsfig{file=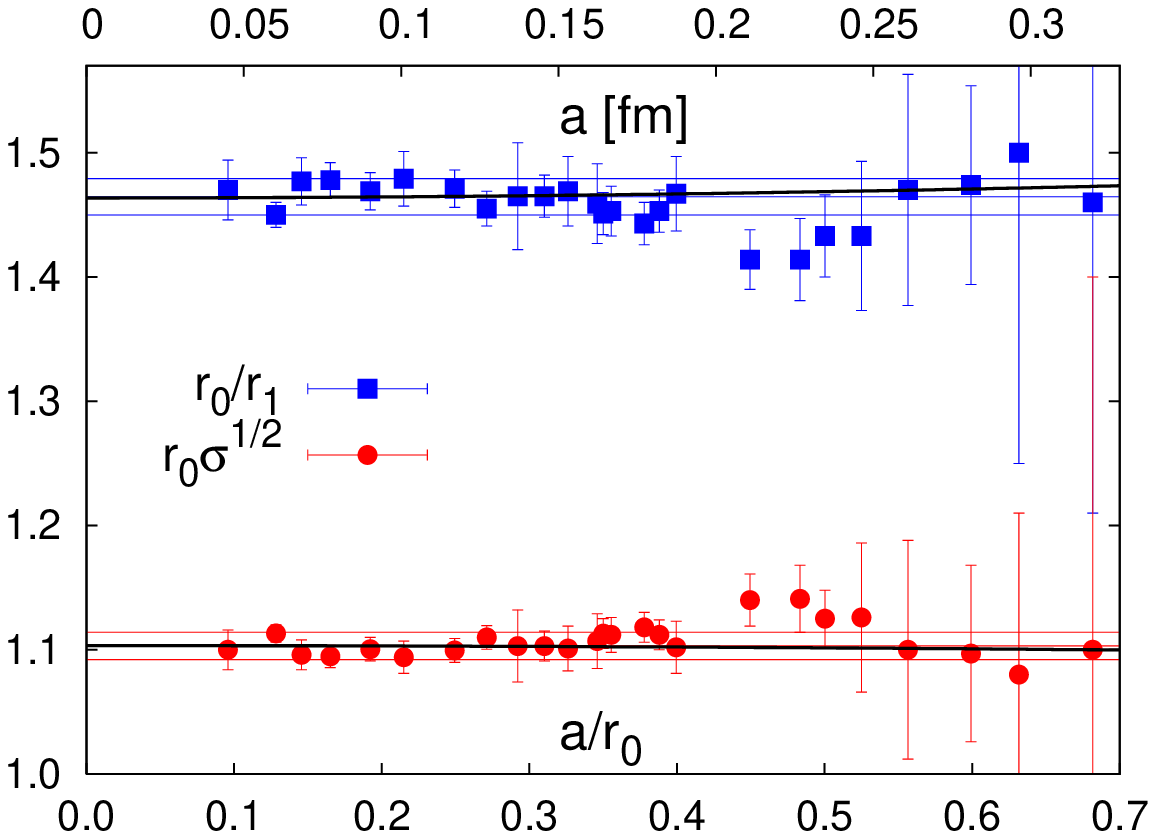, width=8.5cm}
\end{center}
\end{minipage}
\end{center}
\caption{The static quark potential in units of the scale $r_0$ versus 
distance $r/r_0$ (left) and dimensionless combinations of the potential 
shape parameters $r_0/r_1$ 
and $r_0\sqrt{\sigma}$ extracted from fits to these potentials (right). 
The left hand figure shows potentials for several values of $\beta$
taken from our entire simulation interval, $\beta \in [3.15:4.08]$. 
The lowest curve in this figure combines all potentials by matching
them to the string potential (solid line) as explained in the text.
Curves in the right hand figure show quadratic fits and
a fit to a constant with a 1\% error band.
The lattice spacing has been converted to physical units using $r_0=0.469$~fm. 
}
\label{fig:scales}
\end{figure}

The left hand part of Fig.~\ref{fig:scales} shows the static quark potential
for several of our parameter sets. We have renormalized these potentials by 
matching\footnote{Further details on the matching of the zero temperature
heavy quark potentials, its application to the renormalization of the
Polyakov loop and finite temperature free energies will be published elsewhere.}
all potentials at a large distance, $r/r_0=1.5$,
to a common value that is taken to be identical to the large distance
string potential, $V_{\rm string} (r) = -\pi/12 r + \sigma r$. The result of
this matching is shown in the lower part of Fig.~\ref{fig:scales}(left) and
the constant shifts needed to obtain these renormalized potentials are listed
in Table~\ref{tab:masses_r0}.
The good matching of all the potential data obtained at different values 
of the cut-off already gives a good idea of the smallness of finite
cut-off effects in this observable. We note that 
this matching procedure provides renormalization constants for the static
quark potential, which we also will use later to renormalize the
Polyakov loop expectation value.

To further analyze the shape of the static quark potential we determined
the scale parameter $r_0/a$ as well as the square root of the string tension
in lattice units, $\sqrt{\sigma}a$. These parameters have been obtained from 
three and four parameter fits. As 
described in \cite{p4_Tc} the latter fit ansatz has been used to estimate 
systematic errors in our analysis of the scale parameters. 

Results for $r_0/a$ and $\sqrt{\sigma}a$ are
given in Table~\ref{tab:masses_r0}. We note that the 
product $r_0\sqrt{\sigma}$ stays constant on the LCP and changes by less
than 2\% in the entire range of couplings $\beta$ in which 
the lattice cut-off changes by a factor 6.
For $a\le 0.15$~fm we used a quadratic fit ansatz,  
$\left( r_0\sqrt{\sigma} \right)_a =  r_0\sqrt{\sigma} + c (a/r_0)^2$, 
to fit 10 data points. The asymptotic value for $r_0\sqrt{\sigma}$
coincides within errors with a simple average over all values of
$\left( r_0\sqrt{\sigma} \right)_a$ in this interval. This confirms
that ${\cal O}(a^2)$ corrections indeed are small for this product.
Similarly we determined the scale parameter $r_1$ frequently used to 
set the scale in calculations performed on finer lattices. Both fits
for $r_0\sqrt{\sigma}$ and $r_0/r_1$ yield $\chi^2/dof\simeq 0.7$.
From this analysis we obtain the parameters characterizing the shape
of the heavy quark potential at masses in the vicinity of the LCP,
\begin{eqnarray}
r_0\sqrt{\sigma} &=& 1.1034(40) \; , \nonumber\\
r_0/r_1 &=& 1.4636(60) \; .
\label{pot_param}
\end{eqnarray}
We note that the result obtained here for $r_0/r_1$ is in good agreement 
with the
corresponding continuum extrapolated value, $r_0/r_1 = 1.474 (7)(18)$,  
determined with the asqtad action from an analysis of the 
quark mass dependence of this ratio at two different values of
the lattice spacing, $a\simeq 0.12$~fm and $a\simeq 0.09$~fm,  respectively
\cite{asqtad_pot}.
We show results for $r_0/r_1$ and $r_0\sqrt{\sigma}$ calculated at 
parameter sets close to the LCP in Fig.~\ref{fig:scales} (right).

Despite the good scaling behavior of dimensionless combinations of scale
parameters deduced from the static potential, one expects, of course, to 
still find substantial deviations from asymptotic scaling relations that are
controlled by universal 2-loop $\beta$-functions. For the scale parameter
$r_0/a$ we parametrize deviations from asymptotic scaling using a rational
function ansatz,
\begin{eqnarray}
\hr_0 \equiv \frac{r_0}{a} = 
\displaystyle{\frac{1+ e_r \hat{a}^2 (\beta) +f_r \hat{a}^4 (\beta)}{a_r 
R_2(\beta) 
\left(1+ b_r \hat{a}^2 (\beta) +c_r \hat{a}^4 (\beta) +d_r \hat{a}^6 (\beta)
\right)}}
\;\;  ,
\label{fit}
\end{eqnarray}
where
\begin{eqnarray}
R_2(\beta)&=& \exp{\left(-\frac{\beta}{12b_0}\right)}
\left(\frac{6 b_0}{\beta}\right)^{-b_1/(2b_0^2)}
\end{eqnarray}
denotes the 2-loop $\beta$-function of QCD for three massless quark
flavors and $\hat{a} (\beta) = R_2(\beta)/R_2(3.4)$.
With this parametrization it is straightforward to calculate the 
$\beta$-function $R_\beta$ entering all basic thermodynamic observables,
\beqn
R_\beta (\beta) =  \frac{r_0}{a} \left(
\frac{{\rm d} r_0/a}{{\rm d}\beta} \right)^{-1}
\;\; .
\label{r0beta}
\eqn 
Furthermore, we need a parametrization of the $\beta$-dependence of 
the bare quark masses to determine the second $\beta$-function entering
the thermodynamic relations, {\it i.e.}  $R_m (\beta)$ defined in 
Eq.~\ref{massfunctions}. For this purpose we use a parametrization of the
product of the bare light quark mass, $\hm_l$ and $\hr_0$ that takes into 
account the anomalous scaling dimension of quark masses \cite{Gasser},
\begin{equation}
\hm_l \hr_0 = a_m\left( 
\frac{12 b_0}{\beta}\right)^{4/9} P(\beta)  \; ,
\label{mr0}
\end{equation}
with $a_m$ being related to the renormalization group invariant
quark mass in units of $r_0$ and 
$P(\beta)$ being a sixth order rational function that parametrizes
deviations from the leading order scaling relation for the bare quark
mass,
\begin{equation}
P(\beta) = \frac{1+ b_m \hat{a}^2 (\beta) +c_m \hat{a}^4 (\beta)
+d_m \hat{a}^6 (\beta)}{1+ 
e_m \hat{a}^2 (\beta) +f_m \hat{a}^4 (\beta)+g_m \hat{a}^6 (\beta)} \; .
\label{poly}
\end{equation}
This ansatz insures that the parametrization for the two
$\beta$-functions as well as the parametrization of their product,
$R_\beta(\beta)R_m(\beta)$, reproduces the universal 2-loop
results given in Eqs.~\ref{betaasym} and \ref{Rm2loop}. 

\begin{figure}[t]
\begin{center}
\begin{minipage}[c]{17.5cm}
\begin{center}
\epsfig{file=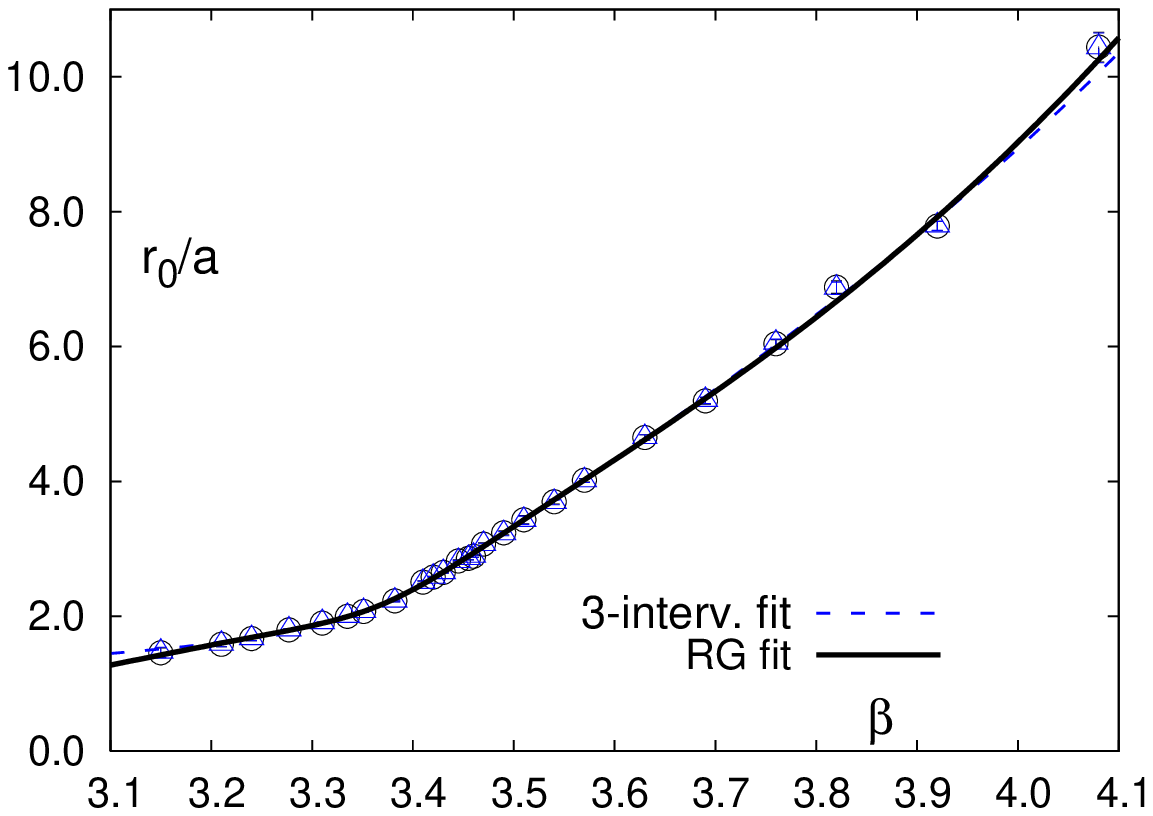, width=8.5cm}
\epsfig{file=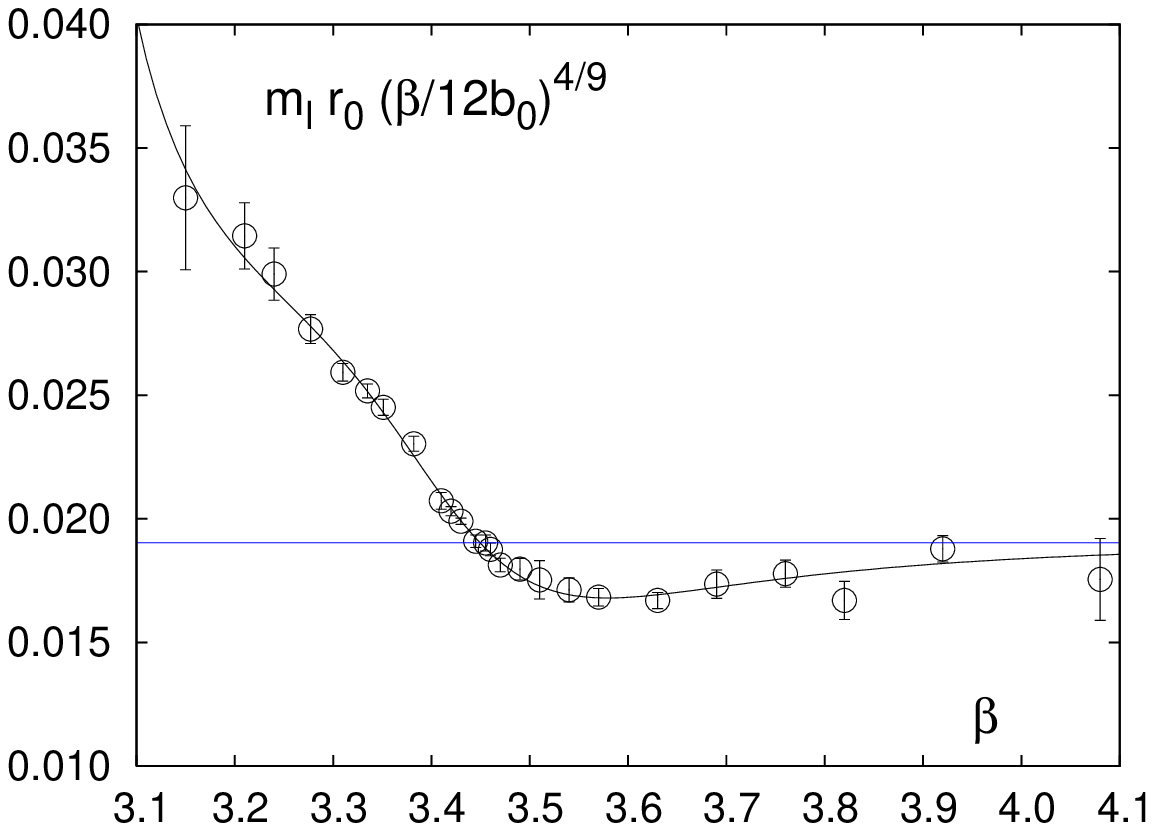, width=8.5cm}
\end{center}
\end{minipage}
\end{center}
\caption{The scale parameter $\hr_0 \equiv r_0/a$ versus $\beta=6/g^2$ (left) 
and its product with the bare light quark mass on the LCP (right).
The two curves shown in the left hand part of this figure correspond to
two different fit ans\"atze. As explained in the text in addition to the 
renormalization group motivated ansatz given in Eq.~\ref{fit} the result 
from a 3-interval fit is shown. The curve in the right hand part of 
the figure shows a fit based on the ansatz given in Eqs.~\ref{mr0} and 
\ref{poly}.
}
\label{fig:LCPfits}
\end{figure}

In Fig.~\ref{fig:LCPfits} we show our results for $\hr_0 = r_0/a$ and 
$\hm_l \hr_0$ together with the fits described above.
The fit parameters defining the quark masses on the LCP have been
obtained from $\chi^2$-fits in the interval $\beta\in [3.1,4.08]$. 
Results for the fit parameters are given in Table~\ref{tab:fits}.
In addition we find 
$a_m = 0.0190(9)$ which turns into a value
of $8.0(4)$~MeV in physical units.
Fit results for $r_0/a$ differ from
the actually calculated values given in Table~\ref{tab:fits} by less than
one percent.

Like in the pure gauge theory calculations of the equation of state,
we also find for QCD with light dynamical quarks that, in the parameter range
of interest for finite temperature calculations, $\beta$-functions
deviate significantly from the asymptotic scaling form.
In particular, we find a 
dip in $R_\beta$ at $\beta \simeq 3.43$. For small values of $N_\tau$, the 
interesting parameter range thus includes the crossover region from the strong 
to weak coupling regime. 

\begin{table}[t]
\begin{center}
\vspace{0.3cm}
\begin{tabular}{|c|c|c|c|c|c|}
\hline
$b_m$ &  $c_m$ & $d_m$ & 
$e_m$ & $f_m$ & $g_m$ \\
\hline
-2.149(121)&1.676(178)&-0.365(144) & 
-2.290(162)&1.829(425)& -0.356(335) \\ 
\hline
\hline
$a_r$ &  $b_r$ & $c_r$ & 
$d_r$ & $e_r$ & $f_r$ \\
\hline
13.250(363)& -1.201(91)& 0.054(196)&
0.406(109)& -1.682(103)& 0.823(76) \\ 
\hline
\end{tabular}
\end{center}
\caption{Parameters of the fit of the scale parameter $r_0$
in lattice units based on the ansatz given in Eq.~\ref{fit} (lower half)
and the fit of  the renormalization group
invariant combination of light quark masses and $r_0$ (Eq.~\ref{mr0}) on the
line of constant physics (upper half).
The $\chi_2/d.o.f$ for these fits are $1.5$ for $19$ degrees of freedom
(lower half) and $0.84$ for $18$ degrees of freedom (upper half). 
}
\label{tab:fits}
\end{table}

We use the interpolating fits for $\hr_0$ and $\hm_l \hr_0$,
to determine the two $\beta$-functions $R_\beta$ and $R_m$ that
enter the calculations of thermodynamic quantities. As all basic
thermodynamic observables are directly proportional to $R_\beta$, we should
check the sensitivity of $R_\beta$ on the particular interpolation form
used. We thus have used a completely different interpolation that
restricts the renormalization group motivated ansatz to the small coupling
regime, $\beta \ge 3.52$, and uses purely rational functions to piecewise
fit 2 intervals at smaller $\beta$. 
We find that results for $R_\beta$ are sensitive to the fit 
ansatz only for small $\beta$-values, {\it i.e} $\beta \lsim 3.25$, 
where the dependence of $\hr_0$ on $\beta$ becomes weak. As discussed
later the uncertainty on $R_\beta$ at small values of the coupling
only affects the three smallest temperatures used for the analysis of the 
equation of state on the $N_\tau=4$ lattices. 

\begin{figure}[t]
\begin{center}
\begin{minipage}[c]{17.5cm}
\begin{center}
\epsfig{file=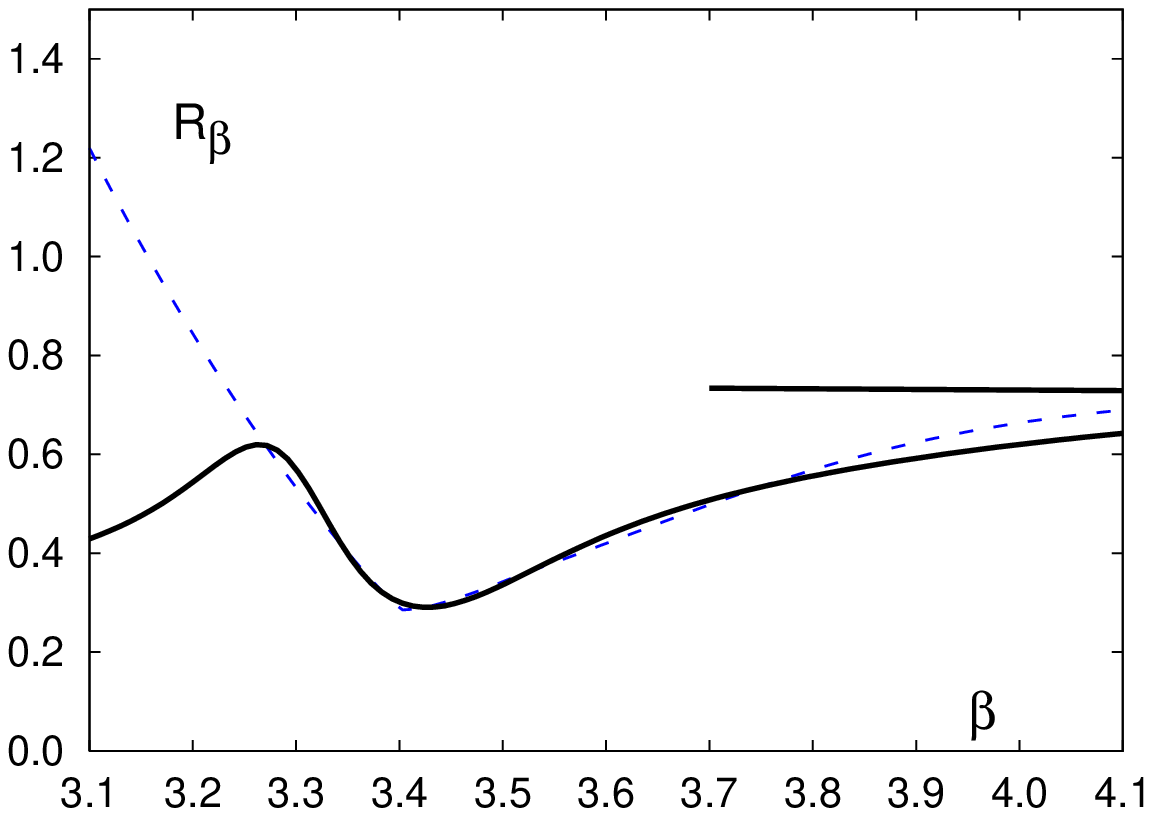, width=8.5cm}
\epsfig{file=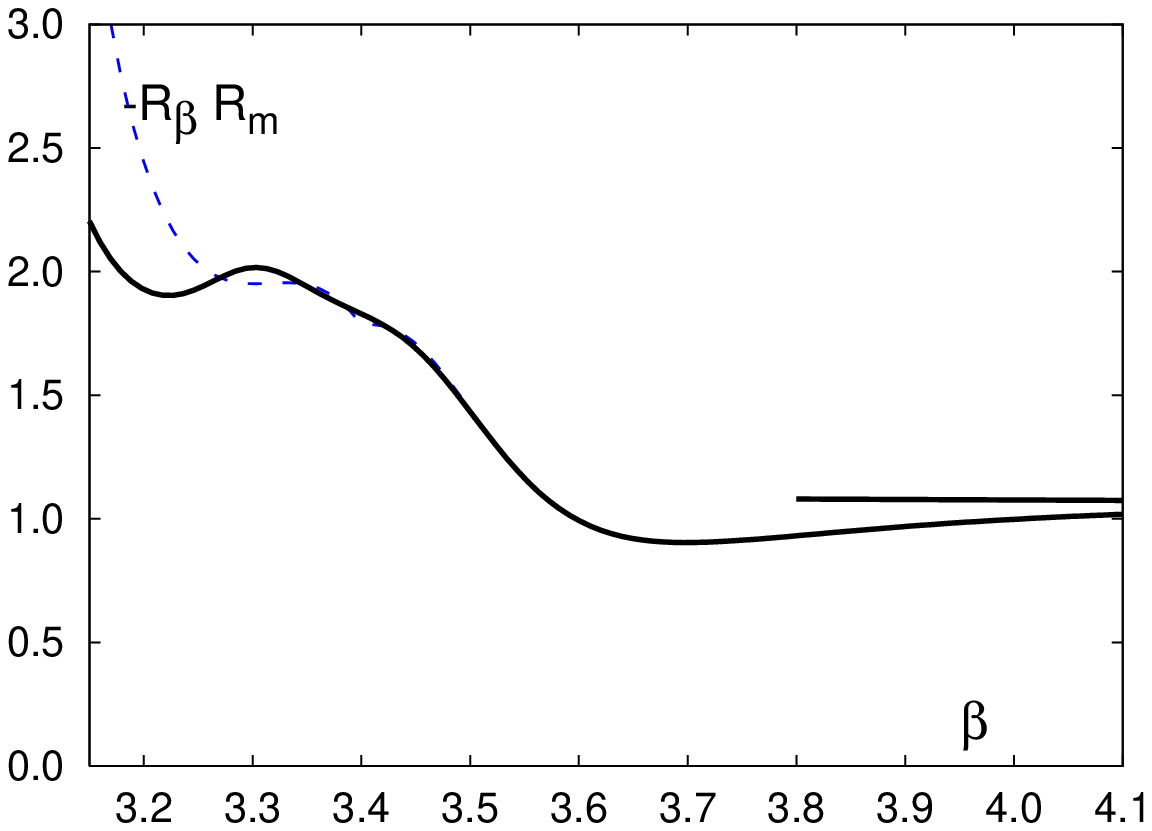, width=8.5cm}
\end{center}
\end{minipage}
\end{center}
\caption{The $\beta$-function on the LCP (Eq.~\ref{r0beta}) (left)
and the product $R_\beta R_m$ (right). The horizontal lines
show the weak coupling behavior given in Eqs.~\ref{betaasym} and \ref{Rm2loop}. 
The two curves result from two different fits of $\hr_0$ as discussed in the 
text. 
}
\label{fig:LCPfits_betafun}
\end{figure}

Using the parametrizations of $\hr_0$ and $\hm_l \hr_0$ given 
in Eq.~\ref{fit} and Eq.~\ref{mr0} as well as the above discussed 
piecewise interpolation of $\hr_0$ 
we now can derive the two $\beta$-functions $R_\beta(\beta)$ 
and $R_m(\beta)$.  
In Fig.~\ref{fig:LCPfits_betafun} we show $R_\beta$ as well as the
combination $-R_\beta R_m$ which enter the calculation of the gluonic and 
fermionic contributions to $(\epsilon - 3p)/T^4$. For $\hr_0$ as well as for 
the two $\beta$-functions, we show result obtained with our two different fit 
ans\"atze.  As can be seen, the different fit forms 
lead to differences in the fit result at the edges of the parameter range 
analyzed. We take care of this in our analysis of 
the equation of state by averaging over the results obtained with the 
two different fit ans\"atze  
and by including the difference of both fit results as a systematic error.
We note that the $\beta$-function $R_\beta$ has a minimum at 
$\beta\simeq 3.43$. This characterizes the
transition from strong to weak coupling regions and is similar to 
what is known from $\beta$-functions determined in pure gauge theory 
\cite{Boyd} as well as in QCD with heavier quark masses \cite{Peikert}. 
The details of this region will differ in  different discretization
schemes as the QCD  $\beta$-functions are universal only up to 
2-loop order in perturbation theory. In order to understand the origin
of cut-off effects in thermodynamic observables it is, however, important
to have good control over $R_\beta$ in this non-universal regime as well, as
$R_\beta$ enters the calculation of all relevant lattice observables as an
overall multiplicative factor\footnote{We note that in order to insure
thermodynamic consistency the $\beta$-function used in the definition of
thermodynamic quantities has to be determined from the cut-off dependence
of the observable used to set the temperature scale, {\it i.e.} $r_0/a$ in
our study.}. 

\section{Bulk thermodynamics}

\subsection{The trace anomaly: \boldmath $(\epsilon -3p)/T^4$}

The basic lattice observables needed to determine the QCD equation of state 
with our tree level improved gauge and fermion actions are expectation values 
of the gauge action as well as the light and strange quark chiral condensates 
calculated on the LCP on finite ($N_\tau \ll N_\sigma$) and zero 
($N_\tau \gsim N_\sigma$) temperature 
lattices. We have performed finite temperature calculations on lattices 
with temporal extent $N_\tau =4$, $6$ and $8$. In all cases the spatial extent
of the lattices ($N_\sigma$) was at least four times larger than the temporal 
extent ($N_\tau$), {\it i.e.} most finite temperature calculations have been 
performed on lattices of size $16^3 4$ and $24^3 6$, respectively. In particular
at high temperature, we found it important to increase the spatial volume in 
our calculations on $N_\tau =6$ lattices to check for possible finite volume effects
and also to add a few calculations on $N_\tau=8$ lattices to get control over the 
cut-off dependence seen in the trace anomaly. In these cases, calculations 
on $32^3 6$ and $32^3 8$ lattices have been performed.
For all parameter sets, corresponding zero temperature calculations have 
been performed on lattices of size $16^3 32$ and $24^3 32$. In a few 
cases we used lattices of size $24^2\cdot 32\cdot 48$ as well as $32^4$.
The length of individual calculations on the finite temperature lattices 
varied between 6500 and 35000 trajectories on the $N_\tau=4$ lattices 
and 5000 to 17600 iterations on the $N_\tau=6$ lattices, where Metropolis 
updates were done after hybrid Monte Carlo evolutions of trajctory length
$\tau_{MD}=0.5$. 
At all values of the gauge couplings the length of runs on zero temperature 
lattices has been adjusted such that the statistical errors of basic 
observables, e.g.
action expectation values, are of similar magnitude as in the $T>0$ runs.
This typically required 2500 to 6000 trajectories. With this amount of
statistics, we achieved statistical errors on the basic thermodynamic
observable, $(\epsilon -3p)/T^4$, of below 20\% at all temperatures. In fact, they
are below 10\% in the temperature
interval $T\in [180{\rm MeV}, 700{\rm MeV}]$ and are less than 5\% for 
$T\in [195{\rm MeV}, 300{\rm MeV}]$. 

\begin{figure}[htb]
\begin{center}
\begin{minipage}[c]{17.5cm}
\begin{center}
\epsfig{file=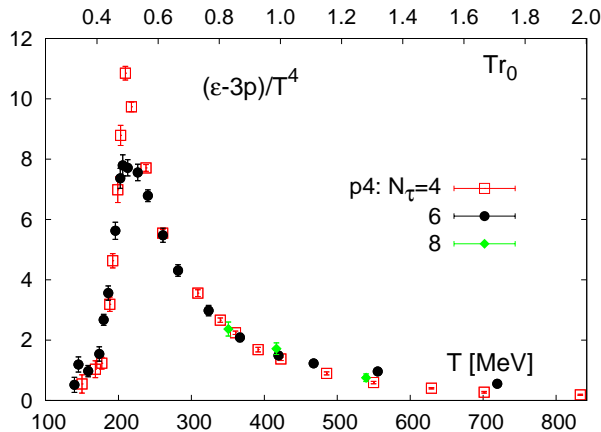, width=8.5cm}
\end{center}
\end{minipage}
\end{center}
\caption{The trace anomaly $\Theta^{\mu\mu}(T) \equiv \epsilon -3p$ 
in units of $T^4$
versus temperature obtained from calculations on lattices with temporal 
extent $N_\tau =4$, $6$, and $8$. 
The temperature scale, $Tr_0$ (upper x-axis) has been obtained using the 
parametrization given in Eq.~\ref{fit}, and $T$~[MeV] (lower x-axis), has
been extracted from this using $r_0=0.469$~fm.
}
\label{fig:e3p}
\end{figure}

The basic zero and finite temperature observables needed to calculate the 
trace anomaly in units of the fourth power of the temperature,
$\Theta^{\mu\mu} (T)/T^4=(\epsilon -3p)/T^4$, 
from Eq.~\ref{deltaLGT} are summarized in 
Tables~\ref{tab:action0}, \ref{tab:action4}, \ref{tab:action6} 
and \ref{tab:nt8}. 
To extract $\Theta^{\mu\mu} (T)/T^4$ one furthermore needs to know
the derivatives of bare couplings and quark masses, 
$R_\beta$ and $R_m$. Their calculation from zero temperature observables
has been discussed in the previous section. 
With this input, we obtain the result for $\Theta^{\mu\mu} (T)/T^4$,
shown in Fig.~\ref{fig:e3p} for the entire range of temperatures 
explored by us.
Here, and in all subsequent figures, the temperature scale has been determined
from our results for $r_0/a$, which characterizes the slope of the
static quark potential and has been extracted from the
zero temperature potential as discussed in the previous section.
On lattices with temporal extent $N_\tau$ we then have 
$T r_0 \equiv \hr_0/ N_\tau$. Whenever we show in the following temperatures 
in units
of MeV we use $r_0=0.469$~fm \cite{gray} to convert $T r_0$ to a MeV-scale. 
We will, however, show in all figures both 
scales which should allow us to compare the results presented here
unambiguously with any other lattice calculation performed within a 
different regularization scheme.
 
\begin{figure}[htb]
\begin{center}
\begin{minipage}[c]{17.5cm}
\begin{center}
\epsfig{file=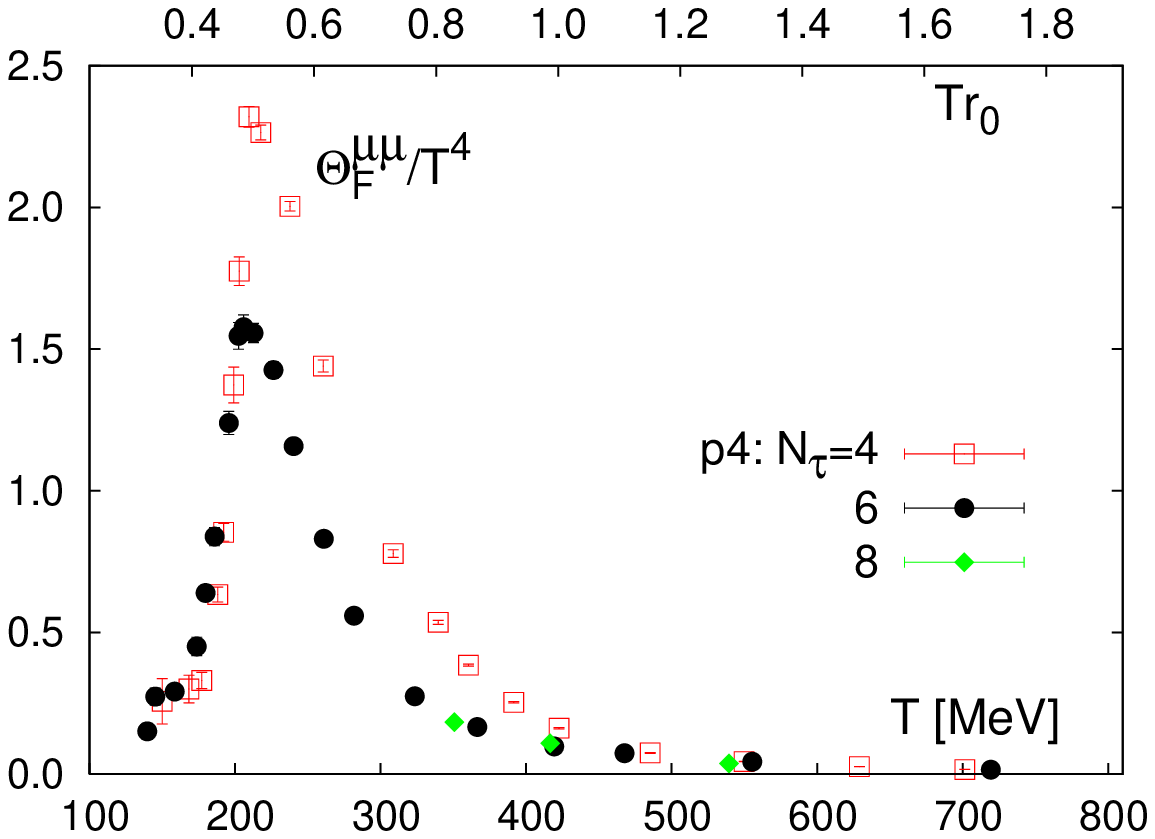, width=8.5cm}
\epsfig{file=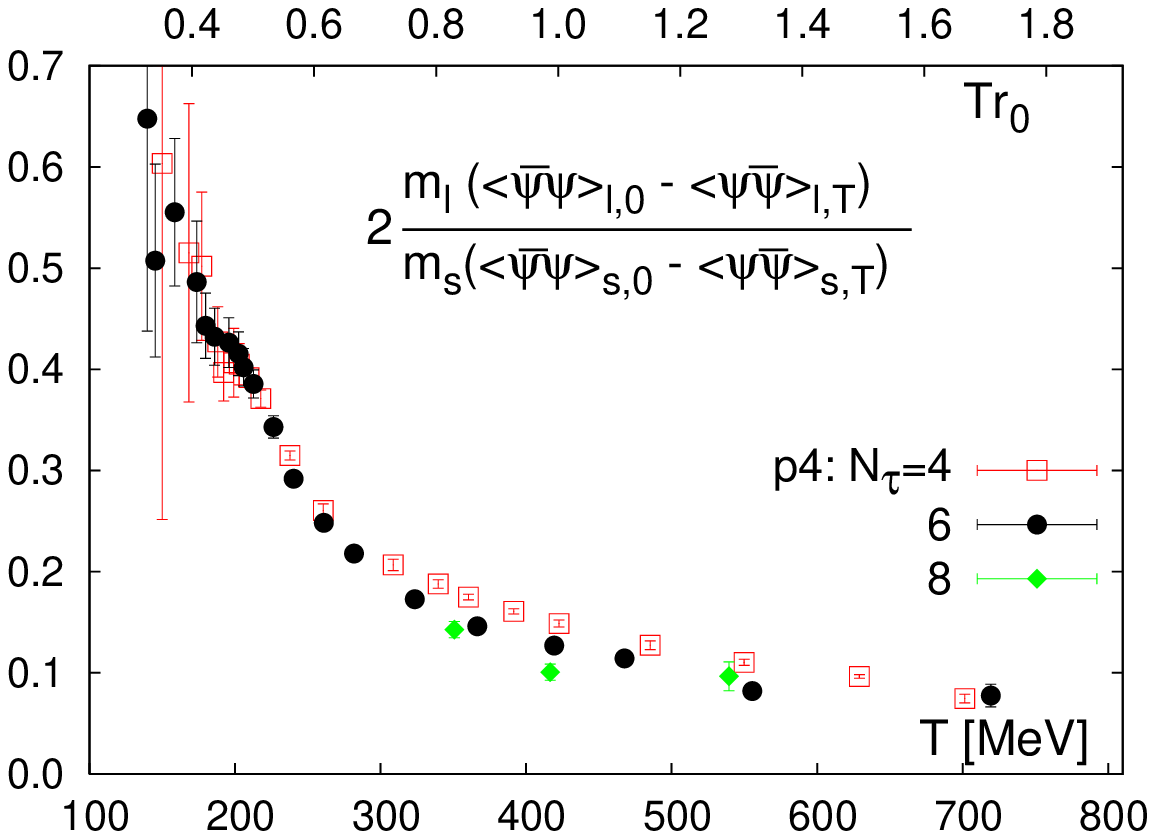, width=8.5cm}
\end{center}
\end{minipage}
\end{center}
\caption{The fermionic contribution to the trace anomaly (left)
and the ratio of the light and strange quark contributions to 
$\Theta_F^{\mu\mu}/T^4$ (right). 
}
\label{fig:e3pF}
\end{figure}

In QCD with light (u, d)-quarks and a heavier strange quark the trace 
anomaly receives, in addition to the gluonic contribution to the trace over 
the energy-momentum tensor, also contributions from the light and heavy quark 
chiral condensates (Eqs.~\ref{e3pgluon}, \ref{e3pfermion}). In the chiral
limit only the former contributes and all fermionic contributions enter
indirectly through modifications of the gauge field background. It thus
is interesting to check the relative importance of direct contributions 
from the chiral condensates to $(\epsilon -3p)/T^4$. In 
Fig.~\ref{fig:e3pF}
we show the fermion contribution $\Theta_F^{\mu \mu}/T^4$ to the total
trace anomaly shown in Fig.~\ref{fig:e3p}. The right hand part of this figure
shows the relative magnitude of the light and strange quark contributions.
As can be seen they are of similar size close to the transition temperature.
With increasing temperature, however, the importance of the light quark
contribution rapidly drops and becomes similar to the ratio of light to
strange quark masses at about twice the transition temperature. As can
be seen in Fig.~\ref{fig:e3pF}(left) the total fermionic contribution shows
a significant cut-off dependence. This partly arises from the large change
of the product of $\beta$-functions, $R_\beta R_m$ that still deviates a 
lot from the asymptotic weak coupling value in the range of couplings relevant 
for the $N_\tau =4$ and $6$ calculations, respectively (see 
Fig.~\ref{fig:LCPfits_betafun}(right)). 
The influence of this cut-off dependence on the calculation of the total
trace anomaly, however, is strongly reduced as the contribution of
$\Theta_F^{\mu \mu}/T^4$ only amounts to about 20\% in the transition region
and already drops below 10\% at about $1.5 T_c$. 

\begin{table}[t]
\begin{center}
\begin{tabular}{|c|c|c|r|l|l|l|l|}
\hline
$\beta$ & $100\hat{m}_l$ & $N_\sigma^3\cdot N_\tau$ &
 $\#~\mbox{traj.}$ &
$\langle s_G \rangle_0$ &
$\langle \bar\psi \psi \rangle_{l,0}$ &
$\langle \bar\psi \psi \rangle_{s,0}$ \\
\hline
 3.150 &    1.100 & $16^3\cdot 32$         &     4544

 &    4.82564(21)
 &    0.28727(11)
 &   0.392677(53)
\\
 3.210 &    1.000 & $16^3\cdot 32$         &     5333

 &    4.68944(27)
 &    0.25284(14)
 &   0.358813(80)
\\
 3.240 &    0.900 & $16^3\cdot 32$         &     5110

 &    4.61441(29)
 &    0.23156(16)
 &   0.333957(88)
\\
 3.277 &    0.765 & $16^3\cdot 32$         &     3408

 &    4.51660(41)
 &    0.20232(17)
 &    0.29834(12)
\\
 3.290 &    0.650 & $16^3\cdot 32$         &     3067

 &    4.47696(37)
 &    0.18807(19)
 &    0.27506(14)
\\
 3.335 &    0.620 & $16^3\cdot 32$         &     3689

 &    4.36044(25)
 &    0.15429(17)
 &    0.24425(10)
\\
 3.351 &    0.591 & $16^3\cdot 32$         &     7005

 &    4.31880(34)
 &    0.14175(20)
 &    0.23045(13)
\\
 3.382 &    0.520 & $16^3\cdot 32$         &     5051

 &    4.23499(26)
 &    0.11515(14)
 &    0.19922(11)
\\
 3.410 &    0.412 & $16^3\cdot 32$         &     5824

 &    4.15990(43)
 &    0.09013(27)
 &    0.16256(20)
\\
 3.420 &    0.390 & $24^3\cdot 32$         &     2448

 &    4.13616(20)
 &    0.08303(17)
 &    0.15304(12)
\\
 3.430 &    0.370 & $24^3\cdot 32$         &     1849

 &    4.11217(29)
 &    0.07606(15)
 &    0.14364(11)
\\
 3.445 &    0.344 & $24^3\cdot 32$         &     1707

 &    4.07770(23)
 &    0.06650(10)
 &   0.130718(86)
\\
 3.455 &    0.329 & $24^3\cdot 32$         &     2453

 &    4.05605(36)
 &    0.06098(24)
 &    0.12314(18)
\\
 3.460 &    0.313 & $16^3\cdot 32$         &     2513

 &    4.04471(35)
 &    0.05733(25)
 &    0.11734(17)
\\
 3.470 &    0.295 & $24^3\cdot 32$         &     3079

 &    4.02346(18)
 &    0.05237(10)
 &   0.109388(88)
\\
 3.490 &    0.290 & $16^3\cdot 32$         &     4300

 &    3.98456(31)
 &    0.04424(22)
 &    0.10072(15)
\\
 3.510 &    0.259 & $16^3\cdot 32$         &     2279

 &    3.94649(29)
 &    0.03657(21)
 &    0.08764(14)
\\
 3.540 &    0.240 & $16^3\cdot 32$         &     4067

 &    3.89302(37)
 &    0.02816(22)
 &    0.07513(17)
\\
 3.570 &    0.212 & $24^3\cdot 32$         &     2400

 &    3.84392(17)
 &   0.021767(89)
 &   0.062829(68)
\\
 3.630 &    0.170 & $24^3\cdot 32$         &     3232

 &    3.75291(10)
 &   0.013176(93)
 &   0.045175(67)
\\
 3.690 &    0.150 & $24^3\cdot 32$         &     2284

 &   3.669908(81)
 &   0.008740(85)
 &   0.035734(47)
\\
 3.760 &    0.130 & $24^2\cdot32\cdot 48$  &     2538

 &   3.580005(77)
 &   0.005781(55)
 &   0.027805(20)
\\
 3.820 &    0.125 & $32^3\cdot 32$         &     2913

 &   3.508124(74)
 &   0.004467(68)
 &   0.024666(37)
\\
 3.920 &    0.110 & $32^3\cdot 32$         &     4677

 &   3.396477(51)
 &   0.002967(69)
 &   0.019635(15)
\\
 4.080 &    0.081 & $32^3\cdot 32$         &     5607

 &   3.234961(31)
 &   0.001546(43)
 &   0.012779(16)
\\
\hline
\end{tabular}
\end{center}
\caption{Expectation values of the pure gauge action density, light and
strange quark chiral condensates calculated on zero temperature lattices
of size $N_\sigma^3N_\tau$. Also given is the number of trajectories 
generated at each value of the gauge coupling $\beta$ with light quarks
of mass $\hm_l$ and bare strange quark mass $\hm_s = 10 \hm_l$.} 
\label{tab:action0}
\end{table}

\begin{table}[t]
\begin{center}
\begin{tabular}{|c|c|l|r|l|l|l|l|c|c|}
\hline
$\beta$ & $100\hat{m}_l$ & $N_\sigma^3$ & $\#~\mbox{traj.}$ &
$\langle s_G \rangle_{\tau}$ &
$\langle \bar\psi \psi \rangle_{l,\tau}$ &
$\langle \bar\psi \psi \rangle_{s,\tau}$ &
$ (\epsilon-3p)/T^4 $ &
$ p/T^4 $ \\
\hline
 3.150 & 1.100 & $16^3$           &    16016
 &    4.82413(46)
 &    0.28165(22)
 &    0.39082(12)
 &       0.54(29)
 &     0.0639
\\
 3.210 & 1.000 & $16^3$           &    21170
 &    4.68525(41)
 &    0.24357(19)
 &    0.35522(12)
 &       1.03(27)
 &     0.1492
\\
 3.240 & 0.900 & $16^3$           &    18741
 &    4.60904(46)
 &    0.21962(26)
 &    0.32920(16)
 &       1.23(18)
 &     0.2060
\\
 3.277 & 0.765 & $16^3$           &    12893
 &     4.5001(12)
 &    0.17784(83)
 &    0.28688(47)
 &       3.18(25)
 &     0.3208
\\
 3.290 & 0.650 & $16^3$           &    30169
 &    4.45142(58)
 &    0.15132(49)
 &    0.25654(28)
 &       4.61(25)
 &     0.4037
\\
 3.335 & 0.620 & $16^3$           &    17327
 &    4.28541(91)
 &    0.04964(84)
 &    0.19082(51)
 &      10.77(20)
 &     1.0757
\\
 3.351 & 0.591 & $16^3$           &    12427
 &     4.2453(11)
 &    0.03744(76)
 &    0.17423(59)
 &       9.68(18)
 &     1.4748
\\
 3.382 & 0.520 & $16^3$           &     8111
 &    4.16623(92)
 &    0.01875(19)
 &    0.13797(43)
 &       7.70(12)
 &     2.2418
\\
 3.410 & 0.412 & $16^3$           &    16000
 &    4.10465(41)
 &   0.011657(41)
 &    0.10229(15)
 &       5.56(12)
 &     2.8435
\\
 3.460 & 0.313 & $16^3$           &    10208
 &    4.00931(64)
 &   0.007148(28)
 &    0.06878(17)
 &       3.57(11)
 &     3.5917
\\
 3.490 & 0.290 & $16^3$           &     9422
 &    3.95941(38)
 &   0.0061563(83)
 &   0.060172(57)
 &      2.668(71)
 &     3.8864
\\
 3.510 & 0.259 & $16^3$           &    10000
 &    3.92564(36)
 &   0.0052568(56)
 &   0.051830(48)
 &      2.249(56)
 &     4.0322
\\
 3.540 & 0.240 & $16^3$           &     6258
 &    3.87812(62)
 &   0.0046270(88)
 &   0.045837(76)
 &      1.687(76)
 &     4.1947
\\
 3.570 & 0.212 & $16^3$           &    21196
 &    3.83212(28)
 &   0.0039044(27)
 &   0.038807(22)
 &      1.378(51)
 &     4.3116
\\
 3.630 & 0.170 & $16^3$           &    10000
 &    3.74581(27)
 &   0.0029122(17)
 &   0.029047(16)
 &      0.896(49)
 &     4.4751
\\
 3.690 & 0.150 & $16^3$           &     7117
 &    3.66559(24)
 &   0.0024312(11)
 &   0.024276(10)
 &      0.592(38)
 &     4.5789
\\
 3.760 & 0.130 & $16^3$           &    33378
 &    3.57727(13)
 &   0.00199846(36)
 &   0.0199662(36)
 &      0.404(22)
 &     4.6498
\\
 3.820(*) & 0.110 & $16^3$           &    32011
 &    3.50620(13)
 &   0.00162776(26)
 &   0.0162683(26)
 &      0.273(28)
 &     4.6830
\\
 3.920 & 0.110 & $32^3$           &     6530
 &   3.395380(89)
 &   0.00154411(10)
 &   0.0154337(10)
 &      0.188(21)
 &     4.7156
\\
\hline
\end{tabular}
\end{center}
\caption{Expectation values of the pure gauge action density, light and
strange quark chiral condensates calculated on lattices with temporal
extent $N_\tau =4$. The last two columns give the trace anomaly, $\epsilon -3p$,
and the pressure, $p$, in units of $T^4$. (*) Note that at $\beta=3.82$
simulations on $N_\tau=4$ and $6$ lattices have been performed at slightly
different quark masses. 
}
\label{tab:action4}
\end{table}

\begin{table}[t]
\begin{center}
\begin{tabular}{|c|c|l|r|l|l|l|l|c|c|}
\hline
$\beta$ & $100\hat{m}_l$ & $N_\sigma^3$ & $\#~\mbox{traj.}$ &
$\langle s_G \rangle_{\tau}$ &
$\langle \bar\psi \psi \rangle_{l,\tau}$ &
$\langle \bar\psi \psi \rangle_{s,\tau}$ &
$ (\epsilon-3p)/T^4 $ &
$ p/T^4 $ \\
\hline
 3.335 & 0.620 & $24^3$           &    14090
 &    4.35980(34)
 &    0.15242(19)
 &    0.24367(13)
 &       0.51(25)
 &     0.0480
\\
 3.351 & 0.591 & $24^3$           &    17610
 &    4.31701(34)
 &    0.13865(20)
 &    0.22923(14)
 &       1.19(25)
 &     0.0686
\\
 3.382 & 0.520 & $24^3$           &    15530
 &    4.23336(35)
 &    0.11103(20)
 &    0.19773(14)
 &       0.97(19)
 &     0.1393
\\
 3.410 & 0.412 & $24^3$           &    10350
 &    4.15710(36)
 &    0.08251(31)
 &    0.15947(19)
 &       1.58(24)
 &     0.2606
\\
 3.420 & 0.390 & $24^3$           &     9550
 &    4.13075(41)
 &    0.07214(39)
 &    0.14812(24)
 &       2.68(19)
 &     0.3347
\\
 3.430 & 0.370 & $24^3$           &    11520
 &    4.10498(50)
 &    0.06110(54)
 &    0.13671(31)
 &       3.57(23)
 &     0.4400
\\
 3.445 & 0.344 & $24^3$           &    14380
 &    4.06634(49)
 &    0.04231(68)
 &    0.11937(35)
 &       5.64(28)
 &     0.6766
\\
 3.455 & 0.329 & $24^3$           &     9050
 &    4.04126(43)
 &    0.02928(64)
 &    0.10788(36)
 &       7.39(32)
 &     0.8982
\\
 3.460 & 0.313 & $24^3$           &     7690
 &    4.02913(42)
 &    0.02374(46)
 &    0.10061(31)
 &       7.82(34)
 &     1.0240
\\
 3.470 & 0.295 & $24^3$           &     9190
 &    4.00834(33)
 &    0.01715(29)
 &    0.09112(24)
 &       7.73(26)
 &     1.2885
\\
 3.490 & 0.290 & $24^3$           &     8360
 &    3.97023(30)
 &    0.01187(19)
 &    0.08185(26)
 &       7.58(27)
 &     1.7784
\\
 3.510 & 0.259 & $24^3$           &     7880
 &    3.93393(23)
 &   0.008204(59)
 &    0.06822(14)
 &       6.80(20)
 &     2.2005
\\
 3.540 & 0.240 & $24^3$           &     6920
 &    3.88347(21)
 &   0.006247(31)
 &    0.05747(15)
 &       5.48(23)
 &     2.7123
\\
 3.570 & 0.212 & $24^3$           &     7310
 &    3.83671(17)
 &   0.004923(12)
 &   0.047364(56)
 &       4.31(19)
 &     3.0925
\\
 3.630 & 0.170 & $24^3$           &     4760
 &    3.74830(17)
 &   0.0034263(61)
 &   0.033892(45)
 &       2.98(17)
 &     3.6137
\\
 3.690 & 0.150 & $24^3$           &     5190
 &    3.66697(15)
 &   0.0027656(24)
 &   0.027530(19)
 &       2.09(14)
 &     3.9362
\\
 3.760 & 0.130 & $24^3$           &     8860
 &    3.57801(12)
 &   0.0022251(10)
 &   0.0222031(93)
 &       1.49(10)
 &     4.1681
\\
 3.820 & 0.125 & $32^3$           &     7870
 &   3.506568(90)
 &   0.00203546(42)
 &   0.0203247(40)
 &       1.23(11)
 &     4.3136
\\
 3.920 & 0.110 & $32^3$           &     9322
 &   3.395328(56)
 &   0.00167642(12)
 &   0.0167504(12)
 &      0.973(86)
 &     4.5057
\\
 4.080 & 0.081 & $32^3$           &     6806
 &   3.234336(54)
 &   0.00114013(10)
 &   0.0113976(10)
 &      0.599(78)
 &     4.7085
\\
\hline
\end{tabular}
\end{center}
\caption{Expectation values of the pure gauge action density, light and
strange quark chiral condensates calculated on lattices with temporal
extent $N_\tau =6$.  The last two columns give the trace anomaly, 
$\epsilon -3p$, and the pressure, $p$, in units of $T^4$.
}
\label{tab:action6}
\end{table}

\begin{table}[t]
\begin{center}
\begin{tabular}{|c|c|c|r|l|l|l|l|c|}
\hline
$\beta$ & $100\hat{m}_l$ &$N_\sigma^3$ & $\#~\mbox{traj.}$ &
$\langle s_G \rangle_{\tau}$ &
$\langle \bar\psi \psi \rangle_{l,\tau}$ &
$\langle \bar\psi \psi \rangle_{s,\tau}$ &
$ (\epsilon-3p)/T^4 $   \\
\hline
 3.820 & 0.125 &  $32^3$ &  15100 
 &   3.507493(335)
 &   0.0021449(31)
 &   0.0213654(187)
 &   2.37(23)
\\
 3.920 & 0.110 &  $32^3$ &   27100 
 &   3.395797(61)
 &   0.00174314(39)
 &   0.0174073(31)
 &   1.72(19)
\\
 4.080 & 0.081 &  $32^3$ &    24100 
 &   3.234705(68)
 &   0.00117543(14)
 &   0.0117488(14)
 &   0.75(13)
\\
\hline
\end{tabular}
\end{center}
\caption{Expectation values of the pure gauge action density, light and
strange quark chiral condensates calculated on lattices with temporal
extent $N_\tau =8$. The last column gives the trace anomaly,
$\epsilon -3p$, in units of $T^4$.
}
\label{tab:nt8}
\end{table}

As all other thermodynamic observables will eventually be deduced from 
$(\epsilon - 3p)/T^4$ using standard thermodynamic relations,
we should analyze its structure carefully. Bulk thermodynamics of QCD in 
different temperature intervals addresses quite different physics.
This includes (i) the low temperature regime, which in the vicinity of
the transition temperature often is compared with the physics of a 
resonance gas and which at lower temperatures is sensitive to properties
of the hadron spectrum controlled by chiral symmetry breaking; (ii) the 
genuine non-perturbative physics in the transition region and at temperatures 
above but close to the crossover
region which is probed experimentally at RHIC and presumably is a still
strongly interacting medium with a complicated quasi-particle structure;
and (iii) the high temperature regime, which eventually becomes 
accessible to resummed perturbative calculations. In numerical calculations
on a lattice these, three regimes also deserve a separate discussion as 
discretization effects influence lattice calculations in these regimes quite
differently. Before proceeding to
a calculation of other bulk thermodynamic observables we therefore will
discuss in the following three subsections properties of $(\epsilon - 3p)/T^4$
in three temperature intervals: 
(i) $T\lsim 200$~MeV or $T\lsim T_c$, 
(ii) $200~{\rm MeV} \lsim T\lsim 300$~MeV  or $1.0\lsim T/T_c \lsim 1.5$ and
(iii) $T\gsim 300$~MeV or $T\gsim 1.5 T_c$.

\subsubsection{Trace anomaly at low temperatures}

\begin{figure}[t]
\begin{center}
\begin{minipage}[c]{14.5cm}
\begin{center}
\epsfig{file=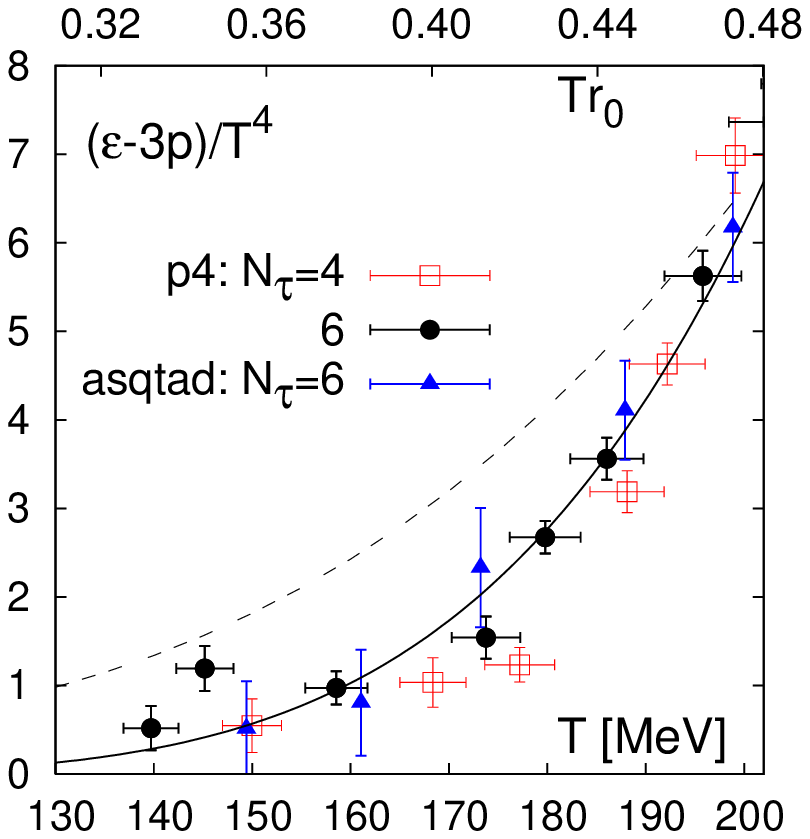, width=12.0cm}
\end{center}
\end{minipage}
\end{center}
\caption{Comparison of the low temperature part of $(\epsilon -3p)/T^4$
calculated on lattices with temporal extent $N_\tau =4$ and $6$
with a resonance gas model that includes all resonances up to mass
$2.5$~GeV (dashed curve). The solid curve shows a polynomial fit to 
the $N_\tau=6$ data obtained with the p4fat3 action. Data
for calculations with the asqtad action are taken from \cite{milc_eos}.
}
\label{fig:e3p_low}
\end{figure}

In Fig.~\ref{fig:e3p_low} we show the low temperature part of 
$(\epsilon-3p)/T^4$ obtained from our calculations with the p4fat3
action on lattices with temporal extent $N_\tau =4$ and $6$ and spatial
size $N_\sigma/N_\tau =4$. We compare these results with calculations
performed with the asqtad action \cite{milc_eos} for $N_\tau =6$. 
These latter calculations have been performed on lattices with smaller spatial
extent, $N_\sigma/N_\tau =2$, and results are based on lower statistics. 
These calculations are, 
however, consistent with our findings. We also note that results obtained
for two different values of the lattice cut-off,
$N_\tau =4$ and $6$, are compatible with each other.

In the transition region from high to low temperature
it is generally expected that thermodynamic 
quantities can be described quite well by a hadron resonance 
gas (HRG) \cite{redlich}; the freeze-out of hadrons in heavy ion 
experiments takes place in this region and observed particle abundances
are, in fact, well described by a HRG model \cite{pbm}. Also 
a comparison of lattice results for the EoS with heavier quarks with
a resonance gas model ansatz was quite satisfactory
\cite{Tawfik} but required the use of a suitably
adjusted hadron mass spectrum.  As we now can perform lattice calculations 
with almost physical quark mass values a more direct comparison
using the HRG model with physical quark mass values
should be appropriate. 

We use a HRG model constructed from all resonances,
with masses taken from the particle data book up to a maximal value
$m_{max} = 2.5$~GeV,
\begin{equation}
\left( \frac{\epsilon - 3p}{T^4}\right)_{low-T} =
\sum_{m_i\le m_{max}}
\frac{d_i}{2\pi^2}
\sum_{k=1}^\infty
(-\eta_i)^{k+1}\frac{1}{k}
\left( \frac{m_i}{T}\right)^3 K_1(km_i/T) \; .
\label{e3plow}
\end{equation}
Here different particle species of mass $m_i$ have degeneracy
factors $d_i$ and $\eta_i = -1 (+1)$ for bosons (fermions).
A comparison of the HRG model with the lattice results is shown
as the upper curve in Fig.~\ref{fig:e3p_low}.

As can be seen in this figure the HRG model captures the qualitative
features of the lattice results on $(\epsilon -3p)/T^4$ quite well, although 
the lattice data seem to drop somewhat faster at low temperature. Whether this
points towards a failure of the HRG model at lower temperatures, or is due
to difficulties in correctly resolving the low energy hadron spectrum in
the current calculations on still rather coarse lattices, will require
more detailed studies on finer ($N_\tau =8$) lattices in the future.
We will return to this question in Section VII.

We also note that
the current lattice calculations are performed with light quark masses that 
are a factor two larger than the physical ones. Reducing the light quark 
masses to their physical
values will shift the lattice data to smaller temperatures and will thus
improve the comparison with the HRG model. From the known systematics of 
the quark mass dependence of other thermodynamic quantities, e.g. the 
transition temperature, chiral condensates, or Polyakov loop expectation
values \cite{p4_Tc} one can estimate this shift to be less than 5~MeV.
Moreover, we note that the scale $r_0$ used to convert lattice results
to physical units has an error of about 2\%. This is indicated in 
Fig.~\ref{fig:e3p_low} by a horizontal error bar for the 
data.
Within this error all data may be shifted coherently.   

The low temperature region of the QCD EoS clearly  
deserves more detailed study in the future.

\subsubsection{Trace anomaly in the strongly non-perturbative regime}

At temperatures just above the transition temperature,
$(\epsilon -3p)/T^4$ shows the largest deviations
from the conformal limit, $\epsilon = 3p$. 
The peak in $(\epsilon-3p)/T^4$  at a temperature $T_{max}$  that is only 
slightly larger than the transition temperature $T_c$ constitutes a prominent 
structure of the trace anomaly which is relatively easy to determine in 
a lattice calculation.
It is closely related to the softest point in the QCD equation of state 
\cite{Shuryak}, {\it i.e.} the 
minimum of $p/\epsilon$ as function of the energy density.
$T_{max}$ thus plays 
an important role for the construction of model equations of state that are
consistent with lattice calculations and may be used in hydrodynamic
models for the expansion of dense matter created in heavy ion collisions.
As $T_{max}$ and, in particular $(\epsilon -3p)/T_{max}^4$, are 
easily determined they may also serve as consistency checks
between different lattice calculations. 

\begin{figure}[t]
\begin{center}
\begin{minipage}[c]{14.5cm}
\epsfig{file=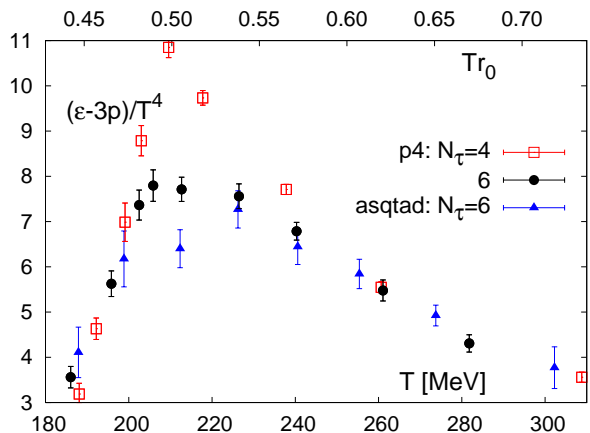, width=8.5cm}
\end{minipage}
\end{center}
\caption{The trace anomaly in the vicinity of the transition temperature 
calculated on lattices with temporal extent $N_\tau =4$ and $6$ on lattices
with aspect ratio $N_\sigma/N_\tau=4$. Data for calculations with the asqtad 
action are taken from \cite{milc_eos} which have been performed on finite
temperature lattices with a smaller physical volume corresponding to
an aspect ratio $N_\sigma/N_\tau=2$. 
}
\label{fig:e-3p_central}
\end{figure}

In Fig.~\ref{fig:e-3p_central} we 
show results for $(\epsilon-3p)/T^4$ in the intermediate temperature 
interval $180~{\rm MeV} < T < 300~{\rm MeV}$. 
Also shown here are results from calculations performed with the 
asqtad action on lattice with temporal extent $N_\tau=6$ \cite{milc_eos}.
As can be seen these calculations are in quite good agreement with the 
results obtained with the p4fat3 action on lattice with the same temporal
extent but larger spatial volume. Estimates for $T_{max}$
and the peak height on lattices with
temporal extent $N_\tau =4$ and $6$ are given in Table~\ref{tab:anomaly}.
Also given there are estimates for
the transition temperature $T_c r_0$ obtained previously in a dedicated
analysis of the transition temperature on $N_\tau=4$ and $6$ lattices 
\cite{p4_Tc}. The values quoted in the Table give the fit results obtained
from a joint fit of transition temperatures on both lattice sizes for
different quark mass values evaluated at the pseudo-scalar mass values
corresponding to our LCP. 
We note that the temperature $T_{\max}$ is
only about 3$\%$ larger than the transition temperatures determined 
from peaks in the chiral susceptibility. 

On the coarse $N_\tau=4$ lattices the
analysis of  $(\epsilon -3p)/T^4$ in the transition region is still quite
sensitive to the non-perturbative structure of the $\beta$-functions,
$R_\beta$ and $R_\beta R_m$ shown in Fig.~\ref{fig:LCPfits_betafun}; 
this region is still close to the strong coupling regime below and in 
the vicinity of the dip in $R_\beta$ shown in 
Fig.~\ref{fig:LCPfits_betafun}(left). This seems to be the main reason
for the large differences seen in the peak height for $(\epsilon -3p)/T^4$
between the $N_\tau=4$ and $6$ lattices. In the latter case the transition
and peak region is already in the regime where the lattice $\beta$-functions
smoothly approach the continuum results. We thus expect that these results
are much less affected by this source of lattice artifacts. Nonetheless,
a better control over the cut-off dependence in this region clearly is 
needed and does require calculations on a larger lattice in order to control 
the continuum extrapolations of $T_{max}r_0$ as well as 
$(\epsilon-3p)/T_{max}^4$. 

\begin{table}[t]
\begin{center}
\vspace{0.3cm}
\begin{tabular}{|c|c|c||c|c|c|}
\hline
$N_\tau$ & $T_c r_0$ & $T_c$ [MeV] & $(Tr_0)_{max}$ & $T_{max}$ [MeV] &$(\epsilon-3p)/T_{max}^4$ \\
\hline
4 & 0.484(4) &204(2)& 0.50(1)&211(4)& 10.8(3) \\
6 & 0.466(6) &196(3)& 0.49(1)&208(4)&  7.8(4) \\
\hline
\end{tabular}
\end{center}
\caption{Position of the peak in $(\epsilon-3p)/T^4$ and its value 
calculated on lattices with different values of the temporal extent
$N_\tau$ on a line of constant physics that corresponds to a pion
mass of about $220$~MeV. Errors on the peak positions have been estimated 
from cubic fits in the peak region by varying the fit intervals. The second 
and third columns show the transition 
temperature determined on the LCP used for this study of the EoS. For 
$N_\tau=4$ this had been determined in \cite{p4_Tc} 
and for $N_\tau =6$ in this analysis (see discussion in Section VI).
}
\label{tab:anomaly}
\end{table}

\subsubsection{Trace anomaly at high temperatures}

\begin{figure}[t]
\begin{center}
\begin{center}
\epsfig{file=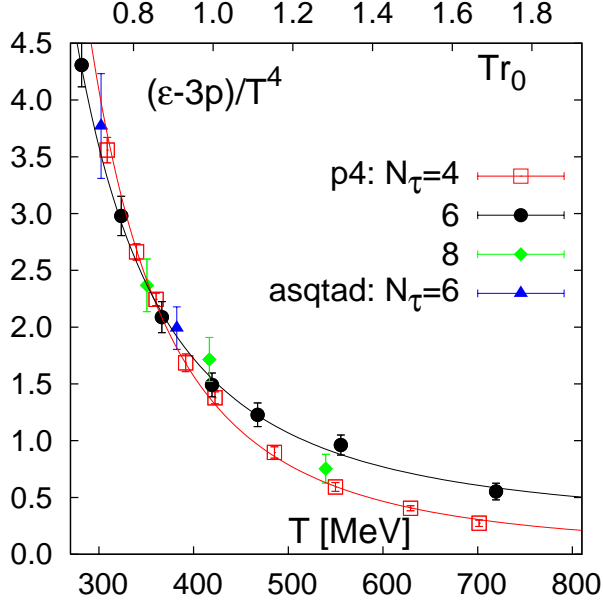, width=12.0cm}
\end{center}
\end{center}
\caption{The high temperature part of $(\epsilon -3p)/T^4$ calculated on
lattices with temporal extent $N_\tau =4$, $6$ and $8$. The curves show fits
to the $N_\tau =4$ and $6$ data with the ansatz given in Eq.~\ref{e3phigh}.
}
\label{fig:e3p_high}
\end{figure}

In Fig.~\ref{fig:e3p_high} we show results for $(\epsilon-3p)/T^4$
in the high temperature regime, $T\gsim 1.5 T_c$. A comparison with
data obtained with the asqtad action on lattices with temporal
extent $N_\tau=6$ shows that the results obtained here with the p4fat3
action are compatible with the former for $T\lsim 400$~MeV ($\sim 2 T_c$). 
The current analysis performed with the p4fat3 action, however, has been 
extended to much larger temperatures, $T\sim 4 T_c$,
{\it i.e.} into the temperature regime accessible to heavy ion experiments
at the LHC.

For temperatures larger than $T_{max}$ the
trace anomaly rapidly drops. Eventually, when the high temperature
perturbative regime is reached, the temperature dependence is 
expected to be controlled by the logarithmic running of the 
QCD coupling constant. To leading
order in high temperature perturbation theory $\Theta^{\mu\mu}(T)$ for
massless quarks is given by \cite{Kap79},
\begin{equation}
{\epsilon -3p \over T^4} = \frac{1}{3} b_0 \left( 1 + \frac{5}{12} n_f
\right) g^4(T)+ O(g^5) \; ,
\label{e3p_pert}
\end{equation}
with $n_f=3$ for massless 3-flavor QCD, which corresponds to the high 
temperature limit for our (2+1)-flavor QCD calculations performed on
a LCP with fixed non-zero quark mass values.

For temperatures larger than about $2.0\; T_c$
results for $\Theta^{\mu\mu}(T)/T^4$ obviously are sensitive to
lattice cut-off effects. The results on $N_\tau=6$ lattices drop 
significantly slower with temperature than the $N_\tau=4$ results. In order
to make sure that this effect does not superimpose with possible finite
volume effects, we increased in this temperature region
the spatial lattice size from $24^3$ to $32^3$. No statistically
significant volume effects have been observed for $\Theta^{\mu\mu}(T)/T^4$,
although we observe a sensitivity of the zero temperature light and strange
quark chiral condensates on the volume; as the condensates contribute less than 
10\% to the trace anomaly at these high temperature values 
(see Fig.~\ref{fig:e3pF}) modifications
of the condensates by a few percent contribute insignificantly to
finite volume effects in $\Theta^{\mu\mu}(T)/T^4$. Moreover, as the entire 
fermionic contribution, $\Theta_F^{\mu\mu}(T)/T^4$, to the total trace anomaly
is small for $T\gsim 400$~MeV, it is obvious that the contribution of the
fermion condensates is not the source for the cut-off effects at high
temperature. The cut-off dependence seen in Fig.~\ref{fig:e3p_high} arises 
from the gluonic sector of $\Theta^{\mu\mu}(T)/T^4$, which of course also
receives contributions from virtual quark loops. 

In the high temperature region we also added calculations on lattices with
temporal extent $N_\tau =8$ at 3 different values of the temperature. 
Results from these calculations are summarized in Table~\ref{tab:nt8} and
are also shown in Fig.~\ref{fig:e3p_high}.
As can be seen in this figure results obtained for the trace anomaly on the 
$N_\tau =8$ lattice are in good agreement with the $N_\tau=6$ results 
suggesting that remaining cut-off effects in this temperature range are small 
for $N_\tau \ge 6$.

We note that larger values for $\Theta^{\mu\mu}(T)/T^4$ at high temperature
also lead to larger values for the pressure, which is obtained from an 
integral over the trace anomaly, and also results in larger values
for the energy and entropy densities, {\it i.e.} these quantities 
approach the Stefan-Boltzmann limit more rapidly on the $N_\tau=6$ 
lattices than they did on the $N_\tau=4$ lattice. It thus is important 
to get good  
control over cut-off effects at high temperatures and obtain further
confirmation of the results obtained in our $N_\tau=6$ lattices, and through
further calculations, on the $N_\tau = 8$ lattices at higher temperatures. 

As discussed previously, $\Theta^{\mu\mu}(T)/T^4$ contains a contribution
from the vacuum quark and gluon condensates that gets suppressed by a 
factor $T^{4}$ at high temperature. In the case of a pure gauge theory it 
has, however, been noted that up to temperatures a few times the 
transition temperature the dominant power-like correction to the 
perturbative high temperature behavior is ${\cal O}(T^{-2})$ rather than
 ${\cal O}(T^{-4})$ \cite{pisarski,Megias}. 
These qualitative features also show up in our results for 
$\Theta^{\mu\mu}(T)/T^4$ at temperatures $T\gsim 1.5T_c$. 
In Fig.~\ref{fig:e3p_high} we show a comparison of the lattice results 
with such a phenomenologically motivated polynomial fit ansatz,

\begin{equation}
\left( \frac{\epsilon - 3p}{T^4}\right)_{high-T} = 
\frac{3}{4}b_0 g^4+\frac{b}{T^2} + \frac{c}{T^4} \; .
\label{e3phigh}
\end{equation}
Here we used the parametric form of the leading order perturbative 
result given in Eq.~\ref{e3p_pert}  with a temperature independent
coupling $g^2$  to characterize
the high temperature behavior of $(\epsilon - 3p)/T^4$
in the fit interval $T\in [300\;{\rm MeV},800\;{\rm MeV}]$. For $N_\tau=4$
we only performed a 2-parameter fit as it turned out that the fit does
not require the contribution from a constant term ($g^2\equiv 0$).
The fit parameters obtained from fits in the region $T\; \ge \; 300$~MeV
are given in Table~\ref{tab:abc_fit}. We note that the 
vacuum condensate contribution ($\sim c/T^4$) is small compared to
the genuine thermal part. The present analysis, however, does not
yet allow us to disentangle logarithmic from power-like (quadratic) 
corrections. 

Of course, it is tempting to
relate the coefficient of the quartic term to a bag constant or zero
temperature quark and gluon condensate contribution, $c= 4B$. 
This yields quite a reasonable value, $B^{1/4} = 247(25)$~MeV. 
Nonetheless, it seems that 
a more detailed analysis of the scaling behavior in the high temperature
region and better control over cut-off effects is needed before a proper
running of the gauge coupling can be established that would unambiguously
allow to single out power-like (quadratic) corrections in the high temperature
regime which then would also allow one to establish a connection to the 
perturbative regime for the trace anomaly.

\begin{table}[t]
\begin{center}
\vspace{0.3cm}
\begin{tabular}{|c|c|c|c|}
\hline
$N_\tau$ & $g^2$ & b [GeV$^2$] & c [GeV$^4$] \\
\hline
4 & -- & 0.101(6) &  0.024(1) \\
6 & 2.3(7) & 0.16(6) & 0.013(6) \\
\hline
\end{tabular}
\end{center}
\caption{Fit parameters for 2-parameter fits ($N_\tau=4$) and 3-parameter
fits ($N_\tau=6$) to $(\epsilon - 3p)/T^4$ in the region $T\;\ge\; 300$~MeV
using the ansatz given in Eq.~\ref{e3phigh}. 
}
\label{tab:abc_fit}
\end{table}

\subsection{Pressure, energy and entropy density}

As indicated in Eq.~\ref{pressure} we obtain the pressure difference,
\begin{equation}
\Delta_ p(T) \equiv \frac{p(T)}{T^4}-\frac{p(T_0)}{T_0^4} \; ,
\label{deltaP}
\end{equation}
by integrating over the trace anomaly 
weighted with an additional factor of $T^{-1}$ in the interval [$T_0,T$].
We have started our integration at $T_0=100$~MeV,
or $Tr_0\simeq 0.24$, by setting the trace anomaly to zero at
this temperature. As discussed in the previous section, this leaves us with
an uncertainty for the value of the pressure at $T_0$, 
which we estimate to be of the order of the 
pressure in a hadron resonance gas, {\it i.e.}
$[p(T_0)/T_0^4]_{HG} = 0.265(2)$. 
The results obtained for $\Delta_p(T)$ 
from our lattice calculations for the pressure at higher temperatures thus 
yield $p/T^4$ up to a systematic uncertainty on $p(T_0)/T_0^4$. 
We also note again that the normalization at $T_0$  does not take care of 
the overall normalization of the pressure at $T=0$. 

\begin{figure}[t]
\begin{center}
\begin{minipage}[c]{17.5cm}
\epsfig{file=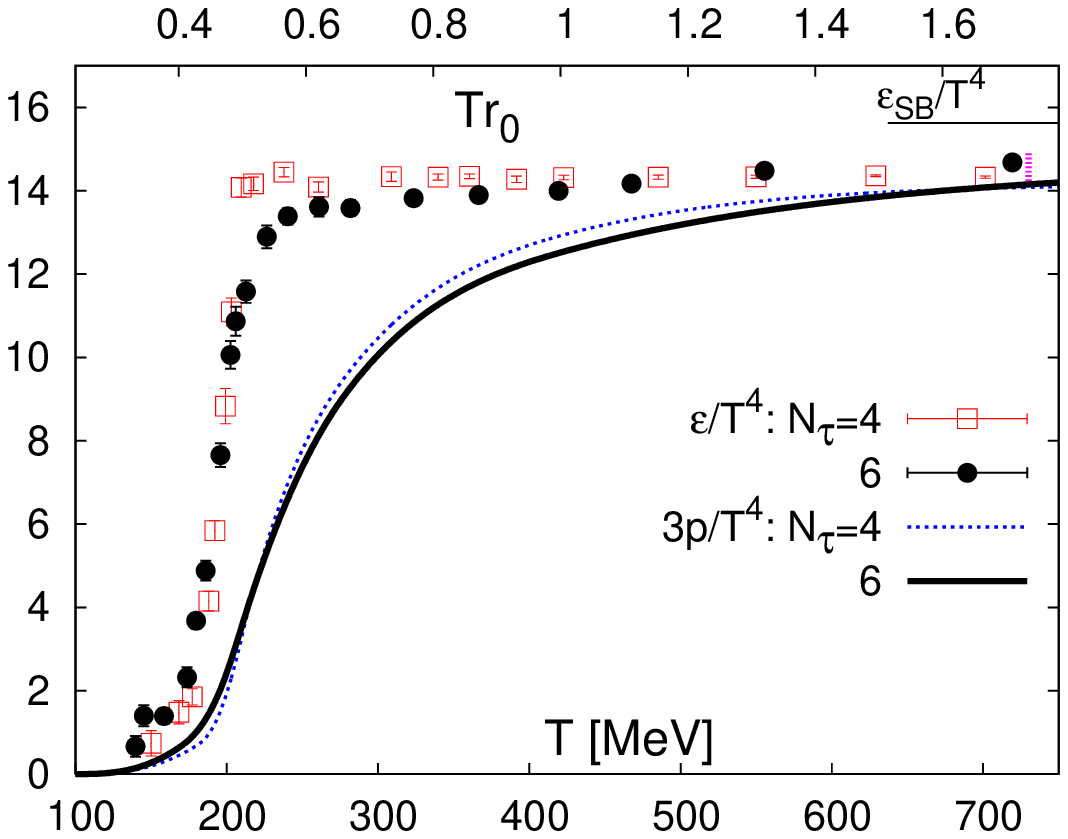, width=8.5cm}
\epsfig{file=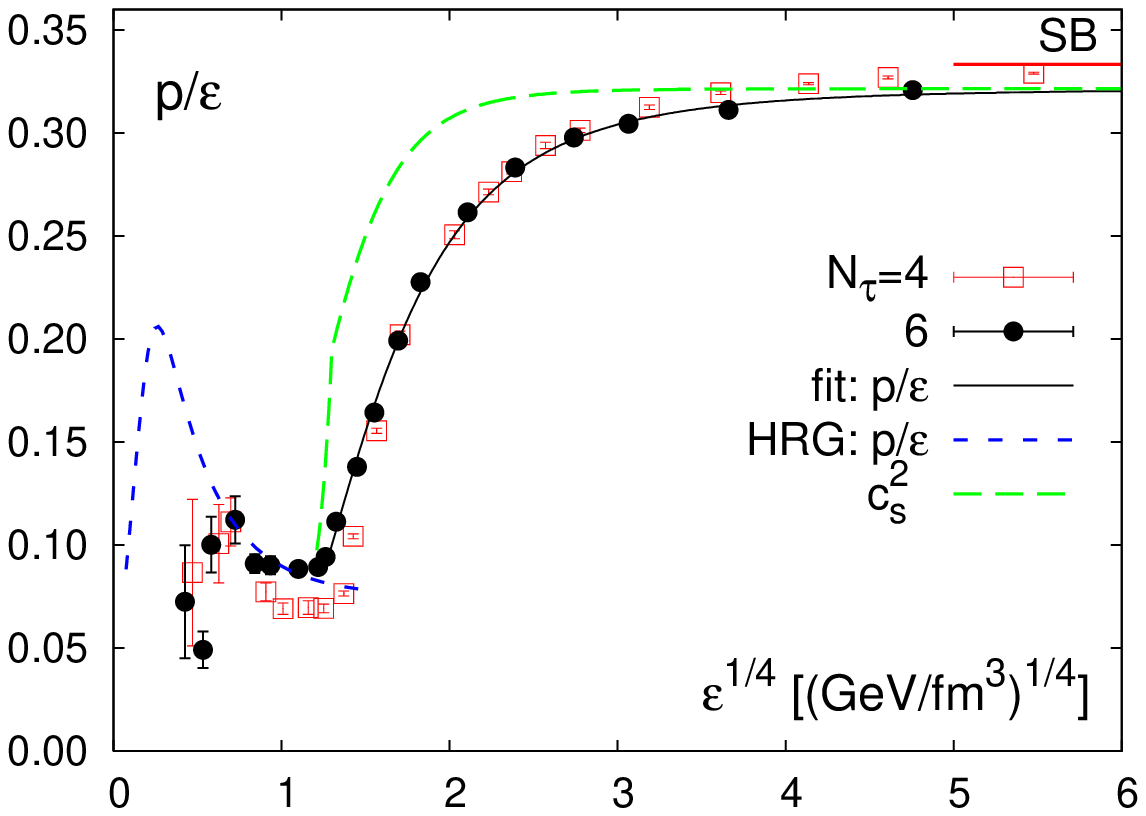, width=8.5cm}
\end{minipage}
\end{center}
\caption{Energy density and three times the pressure as function of the
temperature (left) and the ratio $p/\epsilon$ as function of the fourth
root of the energy density (right) 
obtained from calculations on lattices with temporal extent 
$N_\tau =4$ and $6$. Temperature and energy density scales
have been obtained using the parametrization of $r_0/a$ given
in Eq.~\ref{fit} and $r_0=0.469$~fm. The small vertical bar in the left
hand figure at high temperatures
shows the estimate of the systematic uncertainty on these numbers that
arises from the normalization of the pressure at $T_0=100$~MeV. The dashed
curve (HRG) in the right hand figure shows the result for $p/\epsilon$ in a
hadron resonance gas for temperatures $T < 190$~MeV.
}
\label{fig:eandp}
\end{figure}

To calculate $\Delta_p (T)$ by integrating the numerical results obtained
for $\Theta^{\mu\mu}(T)/T^4$ from Eq.~\ref{pressure}, we have used straight
line interpolations of 
our results for $\Theta^{\mu\mu}/T^4$ at adjacent values of the temperature.
We also used stepwise interpolations obtained by fitting quadratic polynomials 
to the data in small intervals that are matched to fits in the previous
interval. Results of the latter approach are then used to perform the 
integration in the various
regions analytically. Differences between this approach and the straight
line interpolations are nowhere larger than 1.5\%. We then used the smooth
polynomial interpolations to  determine the pressure and combined this
result with that for $\Theta^{\mu\mu} (T)$ to obtain the energy density.
Both are shown in the left hand part of Fig.~\ref{fig:eandp}. The uncertainty
arising from the normalization of the pressure at $T_0$ is indicated
as a small vertical bar in the upper right part of this figure. We note
that at $T \sim 4 T_c$ results for $p/T^4$ and $\epsilon / T^4$ stay about 10\%
below the ideal gas value. 

In particular, for applications to heavy ion phenomenology and for the 
use of the QCD equation of state in hydrodynamic modeling of the expansion
of matter formed in heavy ion collisions, it is of importance to eliminate
the temperature in favor of the energy density and thus obtain the pressure
as function of energy density. The ratio $p/\epsilon$ is shown in the right 
hand part of Fig.~\ref{fig:eandp}. 
As can be seen at low temperature, in the vicinity of the minimum in
$p/\epsilon$, results are consistent with values extracted for this quantity 
from a hadron resonance gas model. We also note that in the high temperature 
regime it has been found in \cite{isentropic} that
the ratio $p/\epsilon$ shows little dependence the baryon number density
when evaluated on lines of constant entropy per baryon number.  

The density dependence of $p/\epsilon$ is related to the square of the
velocity of sound
\begin{equation}
c_s^2 = \frac{{\rm d} p}{{\rm d}\epsilon} = \epsilon 
\frac{{\rm d} p/\epsilon}{{\rm d}\epsilon} + \frac{p}{\epsilon}\; .
\label{sound}
\end{equation}
In the high temperature limit as well as in the transition region where
the derivative ${\rm d} (p/\epsilon)/{\rm d}\epsilon$ vanish, $c_s^2$
is directly given by $p/\epsilon$. We therefore find that the 
velocity of sound is close to the ideal gas value, $c_s^2 = 1/3$,
for energy densities $\epsilon \gsim 100$~GeV/fm$^3$ and drops by a factor
of 4 to a minimal value of about $(c_s^2)_{min} \simeq 0.09$ that is reached
at  $\epsilon \gsim (1-2)$~GeV/fm$^3$. The dependence of $p/\epsilon$ on
the energy density can be parametrized in the high temperature region 
with a simple ansatz \cite{isentropic},
\begin{equation}
\frac{p}{\epsilon} = \frac{1}{3}\left( C - \frac{A}{1+B \epsilon\;
{\rm fm}^3/{\rm GeV}} \right)
\; ,
\label{param}
\end{equation}
which then also allows a simple calculation of the velocity of sound, 
using Eq.~\ref{sound}. We find that the above parametrization yields a 
good fit of the $N_\tau =6$ data in the interval
$1.3\le \epsilon^{1/4}/({\rm GeV} /{\rm fm}^3)^{1/4} \le 6$
with a $\chi^2$/dof of 1.3. For the fit parameters we obtain, $C=0.964(5)$,
$A=1.16(6)$ and $B=0.26(3)$. This fit and the resulting velocity of sound 
are also shown in Fig.~\ref{fig:eandp}(right).

At energy densities below $\epsilon \simeq 1$~GeV/fm$^3$ 
the lattice calculations indicate a rise of $p/\epsilon$
as expected in hadron resonance gas models. However, the current resolution
and accuracy of lattice calculations in this regime clearly is not yet 
sufficient to allow for a detailed comparison between both. 

\begin{figure}[t]
\begin{center}
\begin{minipage}[c]{17.5cm}
\begin{center}
\epsfig{file=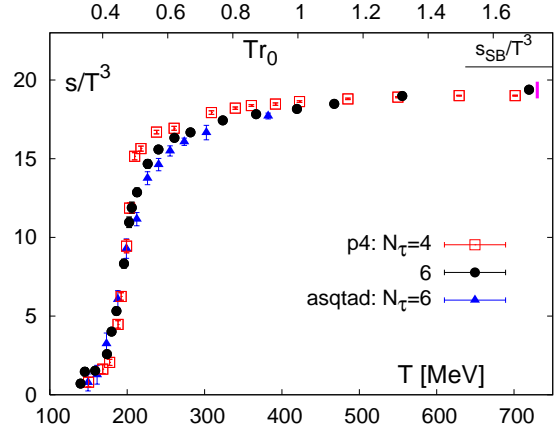, width=8.5cm}
\end{center}
\end{minipage}
\end{center}
\caption{Entropy density as function of the
temperature obtained from calculations on lattices with temporal extent 
$N_\tau =4$ and $6$. Temperature and energy density scales
have been obtained using the parametrization of $r_0/a$ given
in Eq.~\ref{fit} and $r_0=0.469$~fm. The small vertical bar in the left
hand figure at high temperatures
shows the estimate of the systematic uncertainty on these numbers that
arises from the normalization of the pressure at $T_0=100$~MeV. 
}
\label{fig:entropy}
\end{figure}

As pointed out in Section II the non-perturbative vacuum condensates of
QCD show up at high temperature as power-like corrections to temperature
dependence of the trace anomaly and consequently also to
pressure and energy density.
These vacuum condensate contributions drop out in the 
entropy density which is shown in Fig.~\ref{fig:entropy}. It thus is an
observable most suitable for comparisons with (resummed) perturbative
calculations \cite{Blaizot}.
Like energy density and pressure, the entropy also deviates from the ideal
gas value by about 10\% at $T\sim 4T_c$. 

We note that for $T\lsim 2 T_c$ the results obtained with the asqtad 
action \cite{milc_eos} for the entropy 
density are in good agreement with the results obtained with the p4fat3
action, although at least in the high temperature
limit the cut-off dependence of both actions is quite 
different. This suggests that at least up to temperature $T\simeq 2T_c$
non-perturbative contributions dominate the properties of bulk thermodynamic
observables like the entropy density. It also gives rise to the expectation
that additional cut-off effects are small. Nonetheless, 
the result presented in this section on properties of bulk thermodynamic 
observables clearly need to be confirmed by calculations on lattices 
with larger temporal extent.

\section{Renormalized Polyakov loop and chiral condensates}

As part of our analysis of bulk thermodynamic observables we have gathered 
a lot of information on the static quark potential at zero temperature. 
This has been discussed in Section IV and results obtained for 
$V_{\bar{q}q}(r)$ have been used there to determine
a temperature scale for our thermodynamic calculations. Furthermore, 
we have obtained a lot of information on the chiral condensates at zero 
temperature that entered our calculation of thermodynamic quantities.
Together with corresponding results on  heavy quark
free energies and chiral condensates at finite temperature 
this allows us to analyze the deconfining properties as well as the
change of chiral properties of the finite temperature transition
in terms of observables which are related to
exact order parameters for deconfinement and chiral symmetry restoration in 
the infinite quark mass and vanishing quark mass limits of QCD, respectively.

As discussed in the previous sections,
the deconfining aspect of the finite temperature transition, {\it i.e.}
the sudden liberation of partonic degrees of freedom in QCD, is
reflected in the rapid change of bulk thermodynamic observables.
This is also reflected in the rapid change of the static quark free energy 
which characterizes the response of a thermal medium to the addition
of static quark sources.
The static quark free energy, $F_q$, is related 
to the Polyakov loop expectation value, 
$\langle L \rangle \sim \exp (-F_q(T)/T)$,
\begin{equation}
 \langle L \rangle = \left\langle \frac{1}{N_\sigma^3}\sum_{\vec{x}}
L_{\vec{x}} \right\rangle \;\; {\rm with}\;\;
L_{\vec{x}}  = \frac{1}{3}{\rm Tr}
\prod_{x_0=1}^{N_\tau} U_{(x_0,\vec{x}),\hat{0}}  \; .
\label{polyakovx}
\end{equation}
It may more rigorously be defined through the asymptotic large distance 
behavior of static quark-antiquark correlation functions \cite{Zantow}, 
\begin{equation}
\langle L\rangle^2 = \lim_{|\vec{x}-\vec{y}|\rightarrow \infty}
\langle L_{\vec{x}} L_{\vec{y}}^{\dagger} \rangle \; .
\label{polcor}
\end{equation}
The Polyakov loop needs to be renormalized in order to attain a physically 
meaningful value in the continuum limit.  To construct the renormalized 
Polyakov loop from the bare Polyakov loop expectation values,
$\langle L \rangle$, calculated on lattices with temporal extent $N_\tau$ 
at a temperature controlled by the gauge coupling $\beta$,
\begin{equation}
L_{ren}(T) = Z_{ren}^{N_\tau}(\beta) \langle L \rangle \; ,
\label{Lren}
\end{equation}
we can make use of our extensive calculations of the static potential at
zero temperature. As  outlined in Section IV we have extracted renormalization 
constants, $(c(\beta)a)$, from the matching of the static potential to the 
string potential. These renormalization constants are given in 
Table~\ref{tab:masses_r0} in terms of the product $c(\beta) r_0$. 
With this we obtain the renormalization constants for the Polyakov
loop as, $Z_{ren}(\beta)=\exp (c(\beta)a/2)$. 

\begin{figure}[t]
\begin{center}
\begin{minipage}[c]{17.5cm}
\epsfig{file=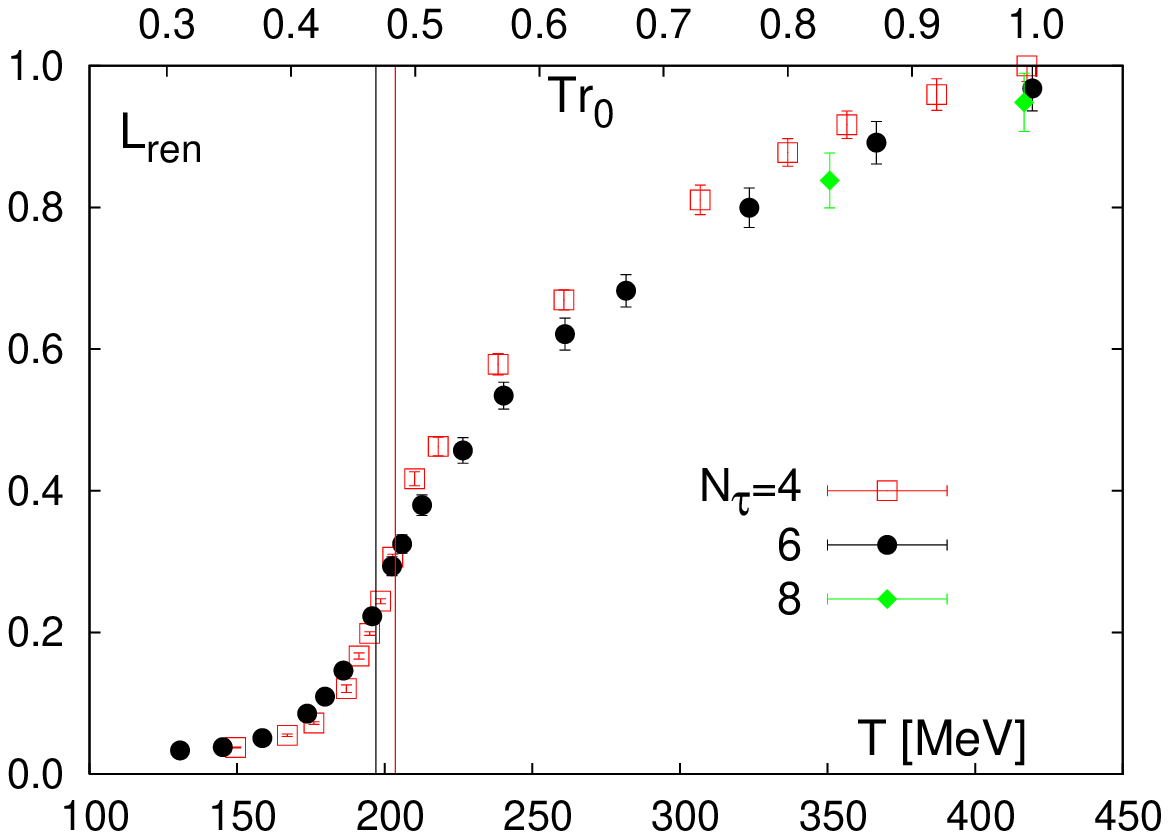, width=8.5cm}
\epsfig{file=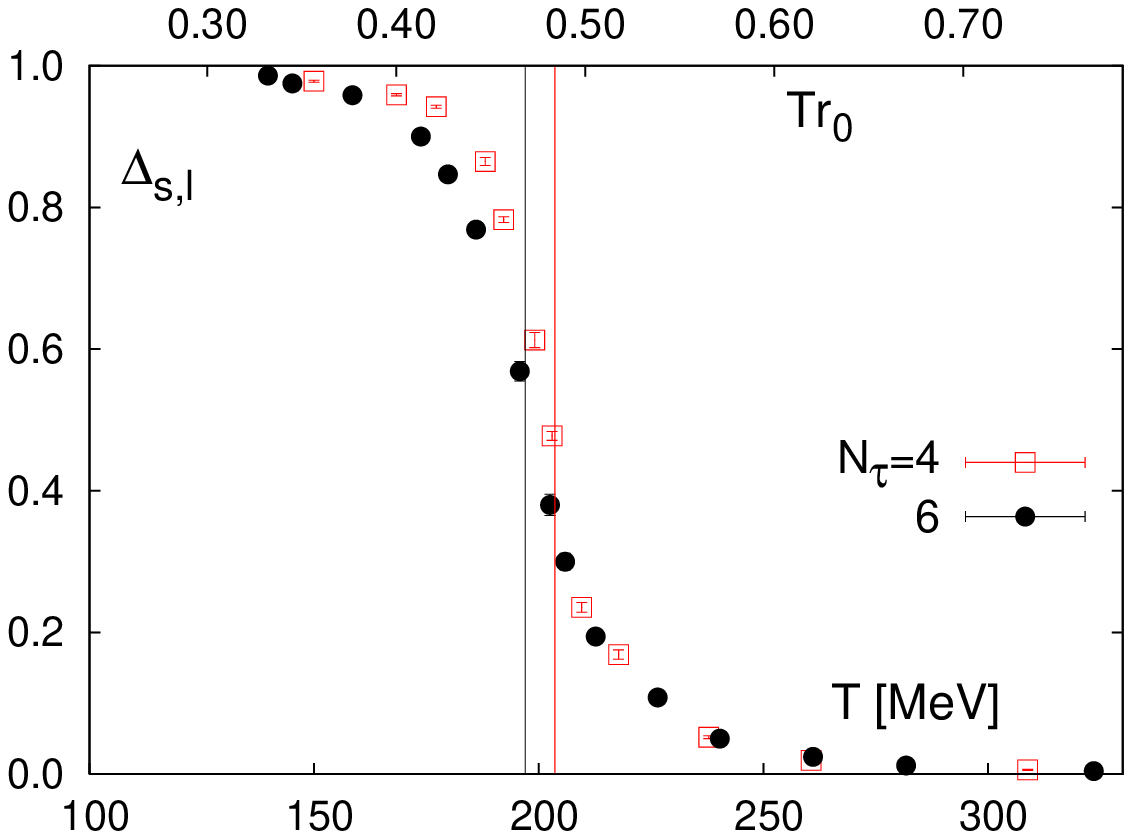, width=8.5cm}
\end{minipage}
\end{center}
\caption{Renormalized Polyakov loop on lattices with temporal extent
$N_\tau=4$, $6$ and $8$ (left) and the normalized difference of light and
strange quark chiral condensates defined in Eq.~\ref{deltapbp}. 
The vertical lines show the location of the
transition temperature  determined in \cite{p4_Tc} 
on lattices with temporal extent  $N_\tau=4$ (right line)
and  in this analysis for $N_\tau=6$ (left line).
}
\label{fig:order}
\end{figure}

Results for the renormalized Polyakov loop are shown in 
Fig.~\ref{fig:order}(left). We note that the cut-off dependence of $L_{ren}$
on lattices with temporal extent $N_\tau=4$ and $6$ is small, which is in 
agreement with results obtained in studies of $L_{ren}$ in pure $SU(3)$
gauge theories \cite{Zantow}. A similar renormalization of the Polyakov
loop obtained in calculations with the 1-link, stout smeared staggered 
fermion action has been used in \cite{fodor}. 
The large cut-off dependence of $L_{ren}$ observed in this case mainly
seems to be due to the choice of observable ($f_K$) that has been used
to set the temperature scale. In fact, when using $r_0$ instead of
$f_K$ to determine the lattice spacing, and thus the temperature, most
of the cut-off dependence of $L_{ren}$ is removed in the data shown in
\cite{fodor}.

Another important aspect of the QCD transition is, of course, the change of 
chiral properties with temperature. This is generally reflected in the 
temperature
dependence of the chiral condensate or related susceptibilities. Also
the chiral condensates need to be renormalized to obtain finite, well
defined quantities in the continuum limit. To eliminate the quadratic 
divergences in the linear quark mass dependent correction to the chiral 
condensates \cite{Gasser} we calculate a suitable
combination of light and strange quark condensates at finite temperature.
We furthermore  
normalize this quantity by the corresponding combination of condensates 
calculated at zero temperature at the same value of the lattice cut-off, 
{\it i.e.} at the same value of the gauge coupling $\beta$,
\begin{equation}
\Delta_{l,s}(T) = \frac{\langle \bar{\psi}\psi \rangle_{l,\tau} -
\frac{\hm_l}{\hm_s}
\langle \bar{\psi}\psi \rangle_{s,\tau}}{\langle \bar{\psi}\psi \rangle_{l,0} -
\frac{\hm_l}{\hm_s} \langle \bar{\psi}\psi \rangle_{s,0}} \; .
\label{deltapbp}
\end{equation}
This eliminates multiplicative renormalization factors. 

In Fig.~\ref{fig:order}(right) we show results obtained for $\Delta_{l,s}(T)$ 
on the LCP for $N_\tau =4$ and $6$. 
In this figure, as well as in the corresponding figure for $L_{\rm ren} (T)$
shown on the left hand side, we also give 
estimates for the pseudo-critical temperature extracted from the  
position of the peak in the disconnected part of the light quark chiral 
susceptibility. For $N_\tau =4$ this value has been determined previously 
by us \cite{p4_Tc} as the quark mass parameters
for the LCP used here are close to those used in \cite{p4_Tc} to determine
$T_c$ on the $N_\tau =4$ lattices for $\hm_l/\hm_s =0.1$. 
For $N_\tau =6$ the choice of the strange quark mass differs slightly
from the one used in that earlier study. We therefore performed a new 
determination of the transition temperature for the $N_\tau =6$ lattice
and the parameters of the LCP used here. From the peak positions of 
the disconnected parts of the light and strange quark susceptibilities
we find $\beta_c(N_\tau=6) = 3.445(3)$. Using the value for $r_0/a$ quoted
for this value of the coupling in Table~\ref{tab:masses_r0} we 
find\footnote{Our
earlier analysis for $m_l=0.1m_s$ on a $16^3 6$ lattice, performed with a 20\%
larger strange quark mass, gave $T_cr_0 = 0.4768(51)$ or $T_c = 201(2)$.}  
$T_cr_0 = 0.466(6)$ or $T_c = 196 (3)$.

We note that the region of most rapid change in the subtracted and normalized
chiral condensate, $\Delta_{l,s}(T) $, is 
in good agreement with the region where the Polyakov loop expectation value
as well as bulk thermodynamic quantities, 
e.g. the energy and entropy densities change most rapidly.

\section{Discussion and Conclusions}
 
We have presented here a detailed analysis of the QCD equation of state 
with an almost physical quark mass spectrum. 
The current calculations have been performed with a physical 
strange quark mass value and two degenerate light quark masses that are
about a factor two larger than the physical average light quark mass value.
In a wide temperature range, results
have been obtained on large spatial lattices close to the thermodynamic limit
for two different values of the lattice cut-off,
corresponding to lattices of temporal extent $N_\tau=4$ and $6$. At high
temperatures additional calculations on lattices with temporal extent
$N_\tau =8$ have been performed, which allow us to control apparent cut-off
effects in this temperature range. All finite temperature calculations have 
been supplemented with corresponding zero temperature calculations to perform
necessary vacuum subtractions and to accurately set the temperature scale. 

At high temperature, $T\gsim 2T_c$, bulk thermodynamic observables
such as pressure, energy and entropy density deviate from the continuum 
Stefan-Boltzmann values only by about 10\% and show little cut-off
dependence. This weak cut-off dependence could only be achieved through
the use of ${\cal O}(a^2)$ improved gauge and fermion actions. 
On the other hand, a closer look at the trace anomaly,
$(\epsilon -3p)/T^4$, from which these quantities are derived,
clearly unravels cut-off effects
when comparing results obtained for the $N_\tau=4$ and $6$ lattices; 
for temperatures $T\gsim 2.5 T_c$ or equivalently $T\gsim 500$~MeV results for
$(\epsilon -3p)/T^4$ on the $N_\tau = 4$ lattices are systematically lower
than for $N_\tau = 6$.
Additional calculations performed on $N_\tau =8$ lattices in this
high temperature region are consistent with the results obtained on
$N_\tau = 6$ lattices and thus suggest that cut-off effects are
small on lattice with temporal extent $N_\tau \ge 6$. 
Of course, this should be confirmed through additional calculations on 
lattices with temporal extent $N_\tau =8$ at larger temperatures.
On these fine lattices it also will be interesting to analyze in more
detail the contribution of charm quarks to the equation of state
\cite{laine,charm}.

Getting better control over the temperature dependence of $(\epsilon -3p)/T^4$ 
at high temperature clearly is important when one wants to make contact between
lattice calculations for e.g. the entropy density and high temperature
perturbation theory. Although our present high statistics analysis seems to 
have achieved good control over the cut-off dependence of $(\epsilon -3p)/T^4$
in this high temperature regime, a more extended analysis of the 
temperature dependence on $N_\tau=8$ lattices is still needed to make firm
contact with perturbative or resummed perturbative calculations. 
 
At low temperatures, $T\lsim 200$~MeV, the influence of cut-off effects is
less apparent. We observe that at a given value of the temperature 
results for $(\epsilon -3p)/T^4$ obtained on the $N_\tau=6$ lattice are 
systematically larger than
those obtained on the $N_\tau=4$ lattice. This is, in fact, expected and
is consistent with the cut-off dependence observed in calculations of the
transition temperature on lattices with temporal extent $N_\tau =4$ and $6$
\cite{p4_Tc}.  Also on lattices with temporal extent  
$N_\tau =8$ indications for this to happen have been found in preliminary 
studies of chiral and quark number susceptibilities as well as Polyakov
loop expectation values \cite{lat07}. 
We thus expect that with increasing $N_\tau$, {\it i.e.}
closer to the continuum limit, the region where $(\epsilon -3p)/T^4$ and all
other thermodynamic observables will start to rise rapidly, continues to
shift towards smaller temperatures. 
A further, although smaller shift of the transition region towards
smaller values of the temperature will arise from an extrapolation to
physical quark masses. Judging from the known temperature dependence of
the transition temperature \cite{p4_Tc} and other thermodynamic observables,
like the Polyakov loop expectation value or quark number
susceptibilities \cite{lat07}, this will amount to a shift of the scale
by a few MeV. In fact, extrapolations of the transition temperature in
quark mass and lattice spacing to the physical point have been performed
by several groups for staggered as well as Wilson fermions
\cite{p4_Tc,milc_chi,Bornyakov}. These extrapolations consistently show
that the quark mass dependence of the transition temperature is weak.
We take the quark mass dependence of the transition temperature as indicator
for the shift of the transition region one has to expect in future calculations
on finer lattices with physical values of the quark masses. We also should
stress that current estimates for the bare lattice parameters that
correspond to physical values of the light quark mass, $\hat{m_q}\simeq
0.04 \hat{m_s}$, of course, are based on studies of the spectrum on lattices
with finite cut-off. Eliminating these systematic effects will require
further calculations on finer lattices. This may also improve the
comparison with model calculations of the equation of state in the
low temperature phase of QCD.

\section*{Acknowledgments}
\label{ackn}
This work has been supported in part by contracts DE-AC02-98CH10886
and DE-FG02-92ER40699 with the U.S. Department of Energy,
the Helmholtz Gesellschaft under grant VI-VH-041, the Gesellschaft
f\"ur Schwerionenforschung under grant BILAER and the Deutsche 
Forschungsgemeinschaft under grant GRK 881. Numerical simulations have 
been performed
on the QCDOC computer of the RIKEN-BNL research center, the DOE funded
QCDOC at BNL, the apeNEXT at Bielefeld University and the new BlueGene/L
at the New York Center for Computational Science (NYCCS). We thank the
latter for generous support during the burn-in period of NYBlue.


\end{document}